\documentclass{statsoc}

\usepackage[a4paper]{geometry}
\usepackage{graphicx}
\usepackage[textwidth=8em,textsize=small]{todonotes}
\usepackage{booktabs}
\usepackage{amsmath}
\usepackage{natbib}
\usepackage{hyperref}
\usepackage{bm}
\usepackage{enumerate}
\usepackage{multirow}
\usepackage{setspace}

\newcommand{\vt}{\mathbf}

\graphicspath{ {./Figure/} }

\title[BMC]{Bayesian Matrix Completion for Hypothesis Testing}
\author[Jin {\it et al.}]{Bora Jin$^{1,*}$, David B. Dunson$^{1}$, Julia E. Rager$^{2}$, David M. Reif$^{3}$, Stephanie M. Engel$^{2}$, and Amy H. Herring$^{1}$}
\address{$^{1}$Duke University,
Durham, USA.}
\address{$^{2}$University of North Carolina at Chapel Hill, Chapel Hill, USA.}
\address{$^{3}$North Carolina State University, Raleigh, USA.}
\email{$^{*}$bora.jin@duke.edu}

\begin{document}

%\singlespacing
\setstretch{1.125}

\begin{abstract}
We aim to infer bioactivity of each chemical by assay endpoint combination, addressing sparsity of toxicology data. We propose a Bayesian hierarchical framework which borrows information across different chemicals and assay endpoints, facilitates out-of-sample prediction of activity for chemicals not yet assayed, quantifies uncertainty of predicted activity, and adjusts for multiplicity in hypothesis testing. Furthermore, this paper makes a novel attempt in toxicology to simultaneously model heteroscedastic errors and a nonparametric mean function, leading to a broader definition of activity whose need has been suggested by toxicologists. Real application identifies chemicals most likely active for neurodevelopmental disorders and obesity.
\end{abstract}

%\vspace{-2em}

% Introduction----------------------------------------------------------------

\section{Introduction}
\label{Intro}
Screening and regulating hazardous chemicals is of great importance and urgency especially as massive numbers of new chemicals are introduced every year. The traditional animal or \textit{in vivo} testing paradigms are infeasible due to financial and time constraints \citep{dix2007ToxCast, judson2010vitro}; in addition, it is desirable to minimise animals used in any testing procedure for ethical reasons. Many organisations such as the World Health Organization's Intergovernmental Forum on Chemical Safety (WHO - IFCS), European Chemicals Agency (ECHA), and United States Environmental Protection Agency (EPA) screen chemicals to measure their potential toxicity and develop alternatives to animal testing. 

As a high-throughput screening (HTS) mechanism has been developed based on \textit{in vitro} assays and a large number of chemicals, EPA and ECHA have had opportunities to operate relatively low-cost and rapid chemical screening programs. For instance, Toxicity Forecaster (ToxCast) and Toxicology in the 21st Century (Tox21) from EPA are designed to identify chemicals that likely induce toxicity in humans and prioritise them for further testing \citep{judson2010vitro}, using HTS methods. ECHA also promotes similar approaches expediting chemical risk prioritisation and assessment for the Registration, Evaluation, Authorisation and Restriction of Chemicals (REACH) regulation \citep{echa2017use}. For the rest of this paper, we use the EPA's ToxCast/Tox21 data as a representative of general HTS data. To the authors' knowledge, ToxCast/Tox21 is the largest in size among existing HTS data in toxicology. Furthermore, it has universal coverage of molecules that many regions including Canada, Japan, and European Union as well as the United States have approved for clinical use \citep{tice2013improving}.  

\begin{figure}[htbp]
\centering
\includegraphics[width=0.55\textwidth]{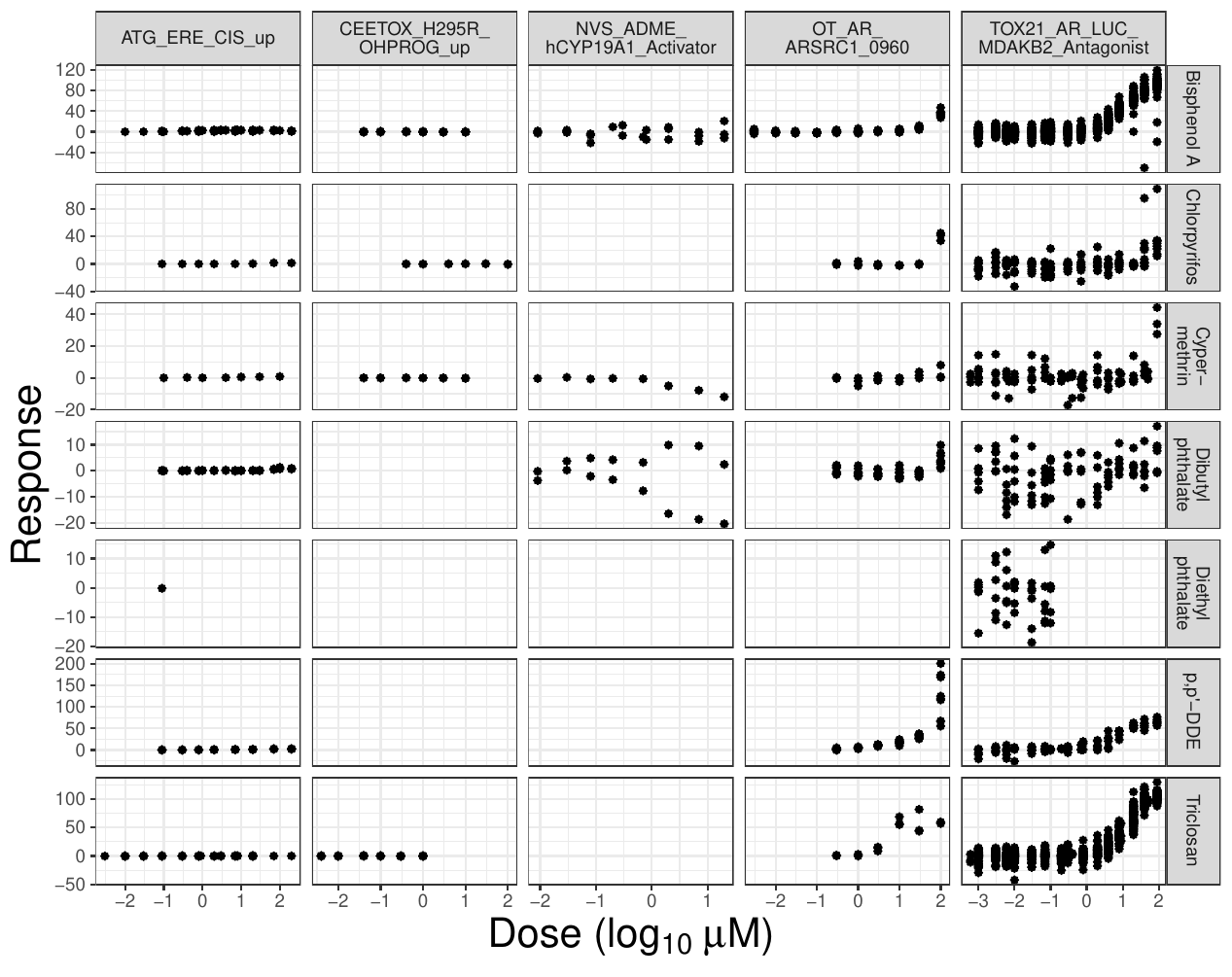}
\caption{Detailed illustration of the ToxCast/Tox21 data structure. Sample data for 7 chemicals on rows and 5 assay endpoints on columns. Each cell contains a set of test results of a single chemical against one assay endpoint, which forms functional data on a dose-response curve.}
\label{DataMat}
\end{figure}

The ToxCast/Tox21 program tests thousands of chemicals against numerous high-throughput assay endpoints. If a chemical exposure leads to biological reactions in an assay, we say the chemical is active for the assay. One of the main goals of these tests is to identify bioactivity profiles of chemicals -- either active or inactive -- across different assays through various endpoints. However, missingness poses a challenge. Although the HTS mechanism has provided a relatively cheap and quick way to conduct millions of tests, it is still only possible to test a small minority of all (chemical, assay endpoint) combinations at irregular doses. This leads to many combinations with few observations or even none as shown in Figure \ref{DataMat} and Figure \ref{DataMatWhole} in the Supporting Material. Figure \ref{DataMat} displays the observed measurements for selected chemicals and assay endpoints from the ToxCast/Tox21 data, and Figure \ref{DataMatWhole} illustrates the overall structure of the data, colour-coded by the number of observations. Here, an observation is the result from an experiment where a chemical is applied to an assay  at a certain dose. We confirm from the figures that the number of observations largely fluctuates across (chemical, assay endpoint) cells; many cells are empty, some have few observations, while others have multiple replicates at a finer grid of doses. 

We arrange the HTS data as matrix-structured functional data, with rows of the matrix corresponding to different chemicals, columns to different assay endpoints, and cells containing sparse dose-response measurements. Our inferential contribution is to construct a complete matrix of the same dimension in which each cell contains a binary indicator for activity using the incomplete and sparse HTS data matrix. Traditional matrix completion focuses on the problem of filling in missing elements of a large matrix based on observations on a small proportion of cells. Typically, the observed cells contain a scalar that is assumed to be measured without error. We are instead faced with a latent binary matrix completion problem. This can be viewed as a matrix-structured multiple hypothesis testing problem for assessing dose-response relationships. 

There have been many Bayesian approaches to multiple hypothesis testing \citep{scott2006exploration, thomas2009use, scott2010bayes, li2010bayesian, scheel2013bayesian, wilson2014hierarchical}. Bayesian approaches are attractive due to their automatic adjustment for multiplicity \citep{scott2006exploration, scott2010bayes} by treating hyperparameters controlling model size as unknown and informed by the data. In the typical framework, hypotheses are considered exchangeable \textit{a priori}. For example, variable selection cases have hypotheses $H_{0j}: \gamma_{j}=0$ and $H_{1j}: \gamma_{j}=1$ in which $\gamma_{j}$ is an indicator of whether the $j$th variable is included for $j=1,\dots,p$, and $\pi_0=Pr(\gamma_j=1)\sim Beta(a,b)$ is a global parameter controlling model size. A variety of more elaborate non-exchangeable priors have been proposed for $\bm{\gamma}=(\gamma_1,\dots,\gamma_p)^T$, designed to include ``prior covariates'' $Z_j$ informing $Pr(\gamma_j=1)$ \citep{thomas2009use} and known structure among covariates represented by an undirected graph \citep{li2010bayesian}. 

There has also been some consideration of matrix-structured multiple testing. In relation to dose-response curves, \cite{wilson2014hierarchical} test for dose effects on the mean using a generalised linear mixed effects model. The mean effect indicator $\gamma_{ij}$ for a (chemical $i$, assay endpoint $j$) pair follows a Bernoulli distribution with $\pi_{ij}=Pr(\gamma_{ij}=1)$. Then $\pi_{ij}$ is further structured with an assay endpoint random effect, chemical-level fixed effect and a probit link: $\pi_{ij} = \Phi(\alpha_j+\alpha x_i)$ where $x_i$ is the chemical-level covariate and $\Phi(\cdot)$ is the cumulative distribution function (CDF) of the standard normal distribution. However, it is nontrivial to find informative chemical-level covariates $x_i$ in this context. In their ToxCast/Tox21 application, \cite{wilson2014hierarchical} found their covariate, chemical solubility, was not significant in explaining $\gamma_{ij}$, resulting in a simplified model with only random effects for assay endpoints. 

In order to account for mechanistic similarities among chemicals and/or assay endpoints as well as to tackle sparsity of the data, we require a more sophisticated hierarchy that borrows information across both rows and columns of the matrix. \cite{tansey2019relational} propose hierarchical functional matrix factorisation methods to infer dose-response curves, approximating the row and the column space using low-dimensional latent attributes. However, their model lacks a formal testing framework. Furthermore, they assume a matrix data structure in which all cells have the same number of replicates at the same number of unique doses, which might not be guaranteed in HTS data. The ToxCast/Tox21 data have different numbers of unique doses within a column and varying numbers of replicates at each dose within a cell. 

We adapt low rank approximations addressing matrix completion problems \citep{mnih2008probabilistic, koren2009matrix, purushotham2012collaborative, tansey2019relational} to a multiple hypothesis testing framework and extend them for more general data structures. This hierarchical Bayesian matrix completion (BMC) approach for hypothesis testing is particularly useful for sparse data. We construct $\pi_{ij}$ with a latent factor model, assuming that low-dimensional latent attributes account for associations relevant to the mean effect among chemicals or assay endpoints. A posterior summary matrix of $\gamma_{ij}$ naturally prioritises chemicals and enables out-of-sample prediction of bioactivity for chemicals not yet tested on certain endpoints, which significantly reduces the amount of \textit{in vitro} testing data that are needed.

Other important characteristics of HTS data are irregular dose-response shapes and heteroscedasticity. Many previous studies placed monotone non-decreasing shape restrictions on dose-response curves \citep{neelon2004bayesian, ritz2010toward, wilson2014hierarchical} and did not consider heteroscedasticity. Our approach is strongly motivated by evidence that disruption in centrality or dispersion of intricately-controlled biological pathways observed \textit{in vitro} can lead to \textit{in vivo} toxicity and ultimately connect to detrimental health effects \citep{klaren2019identifying, knapen2020toward}. Accordingly, a novel attempt in toxicology simultaneously to model heteroscedastic errors as well as any non-constant shapes of the mean completes BMC. This leads to a broader definition of activity as any changes in mean and variance of dose-response curves. These considerations provide a more holistic perspective on active chemicals than previous research. 

The remainder of the paper is organised as follows. Section \ref{Motivations} further explains motivating aspects for a model applicable to HTS data. Section \ref{Data} summarises the ToxCast/Tox21 data of relevance to neurodevelopmental disorders and obesity. Then, the BMC approach is described throughout section \ref{Model}. We compare the performance of BMC with existing methods on simulated data sets and show results using our HTS data in section \ref{Results}, highlighting chemicals that pose greater risks for obesity and neurodevelopment endpoints. Potential areas of future research are discussed in section \ref{Discussion}. The data and the code to reproduce all analyses in the paper are available at \url{https://github.com/jinbora0720/BMC}. 

% Motivations-----------------------------------------------------------------

\section{Motivating Aspects and Relevant Literature} \label{Motivations}
\subsection{Hierarchical Structures}
A simple approach for matrix-structured data would be to consider each cell independently. The EPA has developed an \texttt{R} package \texttt{tcpl} \citep{filer2017tcpl} to facilitate independent dose-response modelling of the ToxCast data. This \texttt{R} package provides three default models: a constant model at zero, a three-parameter Hill model, and a five-parameter gain-loss model for each (chemical, assay endpoint) combination separately. Unfortunately, independently inferring dose-response relationships does not have predictive power: it cannot predict activity for cells having no data. Further, it is likely to have low power and high variance in estimation due to the intrinsic sparsity of the ToxCast/Tox21 data shown in Figures \ref{DataMat} \& \ref{DataMatWhole}. In the ToxCast/Tox21 data, the median number of unique doses tested for each pair is 8, and about 30\% of them are without replicates. Therefore, hierarchical methods for borrowing information are crucial. 

\subsection{Splines without Shape Restrictions}
In estimating dose-response curves, researchers have often forced parametric or monotone restrictions on shapes of the curves to increase interpretability. The EPA's default models currently available through the \texttt{tcpl} package heavily depend on parametric assumptions and are restricted to positive responses to reduce the parameter space, requiring an inverse transformation to fit negative responses. In addition, \cite{wilson2014hierarchical} model dose-response functions by piecewise log-linear splines with constrained parameters to ensure responses are monotone and non-decreasing. In the ToxCast/Tox21 data, it appears difficult to standardise shapes of the dose-response curves (Figure \ref{DataMat}). Furthermore, we observe some examples of decreasing trends between certain assay endpoints (e.g., TOX21\_ERa\_LUC\_BG1\_Agonist as shown in Figure \ref{decreasingfunc} in the Supporting Material) and multiple chemicals. Thus, we propose a non-restricted spline model robust to any shapes of dose-response curves, given that both upturns and downturns in dose-response functions are suggestive of potential toxicity.

\subsection{Heteroscedastic Variances}
Toxicological HTS data have innate heteroscedasticity. Such heteroscedasticity is inevitable because dose effects are variable by nature, with variability often amplified at high doses. Differences in the ability of assays to absorb chemical doses further inflate this variability. \cite{wilson2014hierarchical} attempted to reduce such heteroscedasticity by log transforming the data. However, data transformations may be hard to justify theoretically \citep{leslie2007general} and may be insufficient practically. In genetics, multiple studies have been conducted to detect genetic loci that affect heteroscedastic errors of quantitative traits of interest \citep{pare2010use, ronnegaard2012recent, yang2012fto}. It is widely appreciated that analysing differences in variance could reveal a previously unknown genetic influence and alternative biological relevance. Although detection of heteroscedastic variances is routinely considered in genetic analysis \citep{corty2018vqtl}, it has not been of main interest in chemical toxicity analysis. Without data transformations, we consider heteroscedasticity as another source of information. Fortunately, the ToxCast/Tox21 data have been thoroughly characterised with respect to sources of variability \citep{huang2014profiling, hsieh2015data}, and the performance metrics in \cite{huang2014profiling} indicate that the data have low technical error (e.g. plate-to-plate variation for controls) relative to the heteroscedasticity signals we report. We design an indicator for heteroscedasticity and provide posterior probabilities for the variance effect.

% Data-----------------------------------------------

\section{Data} \label{Data}
This paper uses data from the ToxCast/Tox21 project (invitroDBv3.2, released on March 2019), available at \url{https://epa.figshare.com/articles/ToxCast/Tox21_Database_invitroDB_for_Mac_Users/6062620}. We focus on a subset of the ToxCast/Tox21 data that contain assay endpoints relevant to neurodevelopmental disorders and obesity, along with chemicals tested over those assay endpoints. As a result of selection criteria for chemicals and assay endpoints described in the Supporting Material \ref{s:data}, 30 chemicals evaluated across 131 assay endpoints are studied for neurodevelopmental disorders. These create in total 3930 cells, from which 2024 cells (51.5\%) are missing. For obesity, we use the same 30 chemicals evaluated across 271 assay endpoints. Among the total 8130 cells, 3274 cells (40.3\%) contain no data.

% Model----------------------------------------------

\section{Model} \label{Model}

\subsection{Matrix Completion}
Primary interest lies in differentiating active and inactive chemicals. First, we conduct multiple hypothesis testing of whether the dose-response curve is constant or not across chemicals and assay endpoints. We introduce latent binary indicators $\{ \gamma_{ij} \}$, with $\gamma_{ij}=1$ denoting that the average dose-response curve is not constant for the (chemical $i$, assay endpoint $j$) pair. Let vector $\bm{\gamma}_i=(\gamma_{i1},\dots,\gamma_{iJ})^T$ represent chemical $i$'s {\em mean effect profile} across $J$ assay endpoints for $i=1,\dots,m$. We assume that chemicals and assay endpoints explain dose effects on the mean via low rank latent features, for which we exploit a sparse Bayesian factor model \citep{bhattacharya2011sparse}. Since each $\gamma_{ij}$ takes $\{0,1\}$ values, we impose a generalised factor model using a probit link:
\begin{align}
    Pr(\gamma_{ij}=1) = \pi_{ij} = \Phi(\xi+\bm{\lambda}^T_{i}\bm{\eta}_j) \label{fm}.
\end{align}
A data-augmented form rewrites equation \eqref{fm} as 
\begin{align}\gamma_{ij}=\vt{1}(z_{ij}>0) \text{ where } z_{ij} \sim N(\xi+\bm{\lambda}^T_{i}\bm{\eta}_j,1). \label{fm2}
\end{align}

In the factor model, $\xi$ is an overall intercept, $\lambda_{il}$ represents the coefficient of the $l$th latent pathway for the $i$th chemical to have the mean effect, and $\eta_{lj}$ for the $j$th assay endpoint to have the mean effect for $l=1,\dots,q$ and $q\ll \min(m,J).$ The inequality is reasonable, assuming that not every assay endpoint or chemical forms an idiosyncratic latent pathway for the mean effect. BMC lets either $\bm{\lambda}_i$ be treated as factor loadings and $\bm{\eta}_j$ as latent factors or vice versa, depending on researchers' interests. Provided that one is interested in latent covariance structure among chemicals with regards to the mean effect, a standard factor model puts a multivariate standard normal prior on latent factors $\bm{\eta}_j \sim N_q(\vt{0},I).$ Integrating out $\bm{\eta}_j$ from equation \eqref{fm2} yields $\vt{z}_j \sim N_m(\xi\bm{1}_m, \Lambda\Lambda^T+I)$ where $\Lambda$ has $\bm{\lambda}_i^T$ as its $i$th row. This factor model provides a low dimensional representation of the underlying covariance structure of chemicals.

We employ a multiplicative gamma process shrinkage prior on factor loadings as in \cite{bhattacharya2011sparse}: 
\begin{eqnarray}
\lambda_{il} \sim N(0,\phi^{-1}_{il}\tau^{-1}_l),~\phi_{il} \sim Gamma\left(\nu/2,\nu/2\right),~\tau_l = \prod_{h=1}^l \zeta_h,~l=1,\dots,q, \label{mgsp1}\\ 
\zeta_1 \sim Gamma(a_1,1),~ \zeta_h \sim Gamma(a_2,1),~ h\geq 2. \label{mgsp2}
\end{eqnarray}
This prior choice is supported by \cite{judson2010vitro} who elucidate relationships between chemicals and published pathways: chemicals activate various human genes and pathways, but the number of activated pathways varies widely across chemicals. The multiplicative gamma process shrinkage prior tends to shrink columns of a loading matrix towards zero through the $\tau_l$'s. At the same time, it is possible to strongly shrink only a subset of elements in a certain column through local shrinkage parameters $\phi_{il}$'s, retaining sparse signals. We assume $\xi \sim  N(\mu_{\xi}, \sigma^2_{\xi})$ for the global parameter $\xi$ and recommend to use an informative prior because the information for $\mu_{\xi}$ is largely available from other studies. 

Second, we simultaneously test if heteroscedasticity exists or not. Let $t_{ij} \in \{0,1\}$ indicate changes in the variance of the response with dose, which we denote as variance activity.  Assuming that mean activity and variance activity are likely to be related, we set another factor model for $t_{ij}$ using the factor loadings and the latent factors from the mean activity: $Pr(t_{ij} = 1) = \Phi(\alpha_0 + (\bm{\lambda}_i^T\bm{\eta}_j)\alpha_1)$ in which $\alpha_0$ and $\alpha_1$ are a scalar. Equivalently, 
\begin{align}
    t_{ij} = \bm{1}(u_{ij}>0) \text{ and } u_{ij} \sim N(\alpha_0 + (\bm{\lambda}_i^T\bm{\eta}_j)\alpha_1,1)
\end{align}
Notice that $\alpha_1 > 0$ for positive $\bm{\lambda}_i^T\bm{\eta}_j$ indicates that the $i$th chemical likely to activate the $j$th assay endpoint in the mean is also likely to activate it in variance. If $\alpha_1 = 0$ then heteroscedasticity for each pair is determined by a global parameter $\alpha_0$. The vector $\bm{\alpha} = (\alpha_0, \alpha_1)^T$ has a prior distribution of $N(\bm{\mu}_{\alpha}, V_{\alpha})$.

\subsubsection{Multiplicity Adjustment} 
In our matrix-structured multiple testing, we have $m\times J$ cells, each of which has two hypothesis tests: whether $\gamma_{ij}$ is 0 or 1 and $t_{ij}$ is 0 or 1. Multiplicity is corrected by learning global parameters $\xi$ and $\alpha_0$ from the data. The intuition is that, as $m$ or $J$ increase while the number of true 1's in $\gamma_{ij}$ and $t_{ij}$ remain constant, $\xi$ and $\alpha_0$ will concentrate around large negative values, leading to probabilities near zero after applying $\Phi(\cdot)$. Consider the following posterior distribution for $\xi$ assuming no missing cells,
\begin{align*}
    (\xi|Z,\Lambda, \bm{\eta}) \sim N\left((1/\sigma_{\xi}^2+mJ)^{-1}\left\{\mu_{\xi}/\sigma_{\xi}^2+\sum_{i=1}^m\sum_{j=1}^J(z_{ij}-\bm{\lambda}_i^T\bm{\eta}_j)\right\}, (1/\sigma_{\xi}^2+mJ)^{-1}\right).
\end{align*}
As $m\rightarrow \infty$, $(1/\sigma_{\xi}^2+mJ)\rightarrow \infty$ and $\left\{\mu_{\xi}/\sigma_{\xi}^2+\sum_{i,j}(z_{ij}-\bm{\lambda}_i^T\bm{\eta}_j)\right\}\rightarrow -\infty$ because most of $z_{ij}$ will be negative. Hence, the posterior of $\xi$ will concentrate at large negative values. Similar arguments apply for $\alpha_0$. We show that our model properly adjusts for multiplicity via simulation studies in Section \ref{Simulation}.

\subsection{Dose-Response Functional Data Analysis}
\subsubsection{Splines without Shape Restrictions}
Let $x_{ijk}$ be a test dose (in log base 10 scale in micromolar $(\mu M)$) of the $k$th measurement for a (chemical $i$, assay endpoint $j$) pair, and let $y_{ijk}$ be the corresponding response. Consider the model $y_{ijk} = \gamma_{ij}f_{ij}(x_{ijk})+\epsilon^*_{ijk},$ where the error distribution is $\epsilon^*_{ijk} \sim N(0, \sigma^{*2}_{ijk})$ for $i=1,\dots,m$, $j=1,\dots,J$, and $k=1,\dots,K_{ij}$. Non-constant dose-response curves are estimated when $\gamma_{ij}=1$. We model the dose-response function $f_{ij}$ using cubic B-splines with $p$ degrees of freedom, which is equivalent to estimating $\bm{\beta}_{ij}$ in $(f_{ij}(x_{ij1}), \dots, f_{ij}(x_{ijK_{ij}}))^T = X_{ij}\bm{\beta}_{ij}$ with the B-spline basis matrix $X_{ij}$ of size $(K_{ij} \times p)$. We normalise responses and centre columns of the B-spline basis matrix by $(i,j)$ pairs prior to any analyses in order to exclude the intercept. As Figure \ref{DataMat} suggests, dose-response functions share more similarities within an assay endpoint than between different assay endpoints. This suggests a formulation in which spline coefficients of different chemicals have a common prior covariance matrix for the same assay endpoint. Thus, the prior distributions of spline coefficients and their hyperparameters are $\bm{\beta}_{ij} \stackrel{ind.}{\sim} N_p\left(0,\Sigma_j\right);~ \Sigma^{-1}_j \stackrel{iid}{\sim} Wish_p(a, R^{-1})$ with fixed $a$ and $R$, where $\Omega \sim Wish_p(m, A)$ is a Wishart distribution in $p$-dimensions with $E(\Omega) = mA$. We suggest the following default choices for our application. For assay endpoint-specific covariance matrices, $R$ is determined as the empirical covariance of the ordinary least squares estimates for chemical-assay endpoint pairs. The degrees of freedom parameter $a$ is chosen to be $p+2$ so that $\Sigma_j$ is loosely centred around $R$. While it may seem natural to borrow information also across chemicals, in fact, two chemicals having similar mean activity profiles often have dramatically different dose-response curves. Hence, we are reluctant to borrow information in this manner. 

\subsubsection{Heteroscedastic Variances}
Figure \ref{assayspecificvar} in the Supporting Material illustrates that ranges of responses may vary substantially by assay endpoints. This suggests modelling errors with assay endpoint-specific variances. Moreover, we are motivated to capture heteroscedasticity to explain another dimension of chemical activity. We use a log-linear model on $\sigma^{2*}_{ijk}$ so that $\log \sigma^{*2}_{ijk} = \delta_{0j}+x_{ijk}\delta_{ij}$ and $\sigma^{*}_{ijk} = \exp(\delta_{0j}/2)\exp(x_{ijk}\delta_{ij}/2)$. Here, we separate variance into an assay endpoint-specific variability and a part that changes with dose. Reparametrising $\exp(\delta_{0j}/2)$ with $\sigma_j$ gives the final model equation
\begin{align}
    y_{ijk} = \gamma_{ij}f_{ij}(x_{ijk})+\exp(x_{ijk}\delta_{ij}/2)\epsilon_{ijk}, ~~\epsilon_{ijk} \sim N(0, \sigma_j^2). \label{modeleq}
\end{align}
The assay endpoint-specific variances have an inverse-Gamma distribution \textit{a priori}: $1/\sigma_j^2 \stackrel{iid}{\sim} Gamma\left(\nu_0/2, \nu_0\sigma_0^2/2\right),~\nu_0, \sigma_0^2 \text{ fixed.}$ In our application, we suggest fixing the hyperparameters $\nu_0$ at 1 and $\sigma_0^2$ at the sample variance of the response variable to have the prior distribution weakly centred around a simple estimate from data. 

We assume a spike-and-slab prior for coefficients $\delta_{ij}$ such that $\delta_{ij}\stackrel{iid}{\sim} N(0,v_{\delta})$ if $t_{ij}=1$ with fixed $v_{\delta}$, and $\delta_{ij}=0$ if $t_{ij}=0$. In our case, we found that ensuring a large enough value for $v_{\delta}$ that appropriately covers the data range improves estimation of $\delta_{ij}$ and $t_{ij}$. Provided that conditional standard deviations of responses given doses can be proxies for $\exp(x_{ijk}\delta_{ij}/2)$ in equation \eqref{modeleq}, a range of $\delta_{ij}$'s is obtained. The variance parameter $v_{\delta}$ of $\delta_{ij}$ is then determined as the square of the range divided by 4, which makes $\pm 2$-standard deviation intervals for $\delta_{ij}$ cover its sample range. We finally fix $v_{\delta}$ at the maximum of the above value and the sample variance of the response variable. Combined with $\sigma_0^2$, this allows the prior distributions of two variance parts -- the assay endpoint-specific and the heteroscedastic variance -- to place enough probability on the observed variability from the data.

In conclusion, equation \eqref{modeleq} is the final model in which $\gamma_{ij}$ is an indicator specifying whether the $i$th chemical activates the $j$th assay endpoint in the mean, $t_{ij}$ is an indicator if the $i$th chemical activates the $j$th assay endpoint in variance, $f_{ij}$ is a dose-response function, the exponential term allows for modelling heteroscedastic residual variance, and measurement error is modeled with normal distributions having assay endpoint-specific variances. We suggest a new metric for activity including mean perturbation as well as variance perturbation, which is computed as $\bm{1}(\gamma_{ij}=1 \cup t_{ij}=1)$. 
    
\subsection{Posterior Computation} 
Our posterior samples are obtained using Metropolis-Hastings steps within a partially collapsed Gibbs sampler. Most of the parameters have conjugate posterior distributions which lead to a straightforward update. Details are provided in the Supporting Material \ref{s:posteriors}. The code for the sampler to automate any relevant analyses is readily available to other researchers at \url{https://github.com/jinbora0720/BMC}. 

% Results--------------------------------------------

\section{Results} \label{Results}
\subsection{Simulations} \label{Simulation}
Simulation studies were conducted to evaluate the performance of BMC in learning latent correlation structures among chemicals and  predicting activity probabilities. Two broad scenarios of simulations were examined corresponding to data simulated from BMC (Simulation 1) or an alternative (Simulation 2). For predictive performance, BMC was compared to three variations in the prior structure of $\gamma_{ij}$. Instead of a latent factor model, we assume simpler structures \textit{a priori} as follows: 
\begin{eqnarray}
Pr(\gamma_{ij}=1)=\pi_0 ~\forall i, j, \text{ and } \pi_0 \sim Beta(1,1); \label{bmc0}\\
Pr(\gamma_{ij}=1)=\pi_i ~\forall j, \text{ and } \pi_i \stackrel{iid}{\sim} Beta(1,1)~\forall i; \label{bmci}\\
Pr(\gamma_{ij}=1)=\pi_j~\forall i, \text{ and } \pi_j \stackrel{iid}{\sim} Beta(1,1) ~\forall j. \label{bmcj}
\end{eqnarray}
We call models with equation \eqref{bmc0}, \eqref{bmci} and \eqref{bmcj} BMC$_0$, BMC$_i$ and BMC$_j$, respectively. BMC$_i$ assumes that each chemical has its own intrinsic mean effect probability, while BMC$_j$ assumes that each assay endpoint has its own mean effect probability. These three variations assume a simpler structure for the heteroscedasticity indicator such that $t_{ij}\sim Bernoulli(\pi_t)$ and $\pi_t\sim Beta(1,1)$. For estimation performance, the proposed model is compared to the zero-inflated piecewise log-logistic model (ZIPLL) \citep{wilson2014hierarchical} and tcpl \citep{filer2017tcpl}. The ZIPLL code at \url{https://github.com/AnderWilson/ZIPLL} utilises a Bayesian hierarchical approach whose testing framework for the mean effect adopts equation \eqref{bmc0}. Since the code does not allow missing pairs in the data, we only use ZIPLL for estimation and not prediction. The tcpl models are currently used by EPA and treat dose-response curves independently. 

\begin{figure}[htbp]
\centering
\includegraphics[width=\textwidth]{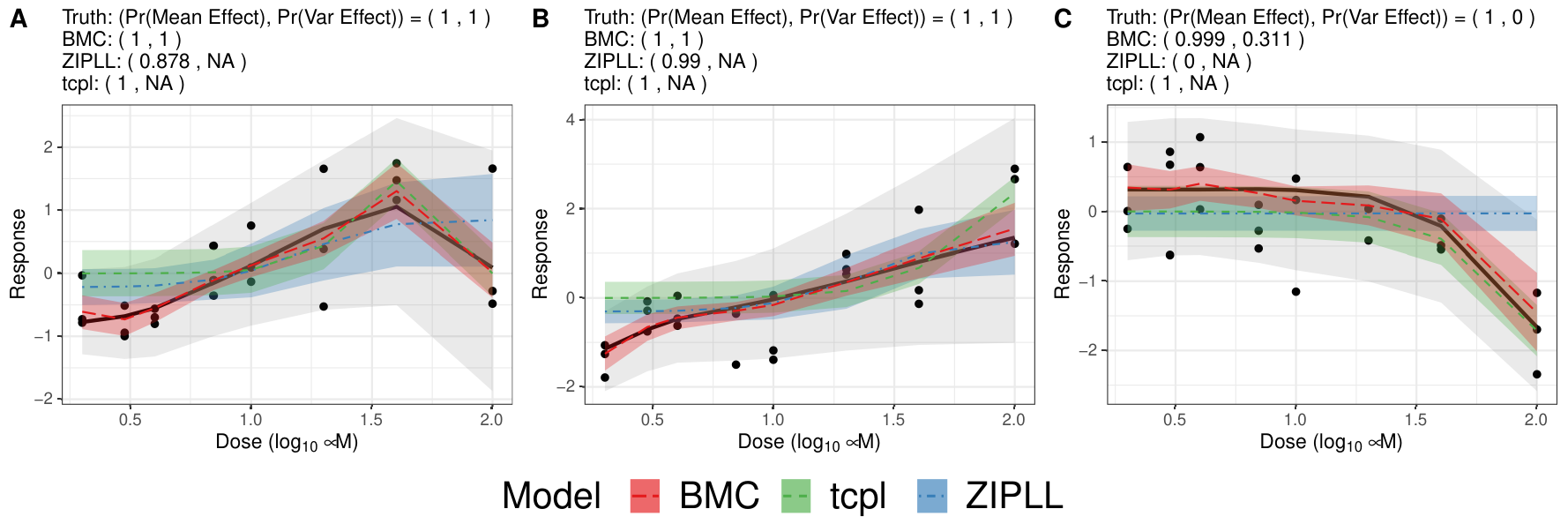}
\caption{Dose-response curves (solid line) with fitted mean functions by BMC (long dash line), ZIPLL (dot-dashed line), and tcpl (dashed line) in Simulation 1. The true curve is mostly increasing and decreasing at higher dose in \textsf{A}, monotonically increasing in \textsf{B}, and monotonically decreasing in \textsf{C}. Shaded areas around estimated functions of the same colour represent 95\% CIs for the average dose-response curves computed by BMC and ZIPLL, and confidence intervals by tcpl. Light gray areas illustrate 95\% posterior predictive intervals from BMC for data points.}
\label{Sim1:curves}
\end{figure}

In Simulation 1 in which BMC is the true data generating process, mimicking the ToxCast/Tox21 application, the number of chemicals $m$ was set to 30, and the number of assay endpoints $J$ to 150. We generate 30 data sets, and in each set we hold out 225 pairs at random, which are 5\% of the total cells in the data matrix. The profiles of the mean effect for chemical-assay endpoint pairs were sampled assuming a factor model, which induced a correlation structure among chemicals (Figure \ref{Sim1:corr} in the Supporting Material). The overall intercept $\xi$ is set at 0. For pairs having dose effects on the mean, dose-response functions were given as one of the three categories: mostly increasing and decreasing at higher doses; monotonically increasing; and decreasing. Figure \ref{Sim1:curves} presents examples of dose-response functions of each category. Heteroscedasticity is assumed to be positively associated with the mean effect, i.e., $\alpha_1=1.2$ with $\alpha_0 = -0.1$. More specific settings of Simulation 1 are described in the Supporting Material \ref{s:sims}. 

As illustrated in Figure \ref{Sim1:curves}, BMC accurately captures true curves regardless of shapes. It also produces tighter 95\% credible intervals (CIs) for the average dose-response curves than competitors. The competitors, ZIPLL and tcpl models, do not seem robust enough to various dose-response curves. In particular, ZIPLL estimates a decreasing trend as constant, which is evident in Figure \ref{Sim1:curves}\textsf{C}. For generally increasing curves (\textsf{A, B}), the ZIPLL and tcpl models sometimes miss the true dose-response functions, which becomes more noticeable when heteroscedasticity exists. These results suggest that in some cases, BMC can lead to more precise inferences on values estimated through dose-response curves, such as Emax (greatest attainable response) or AC50 (chemical dose producing half maximal response in an assay endpoint).

\begin{figure}[htbp]
\centering
\includegraphics[width=0.8\textwidth]{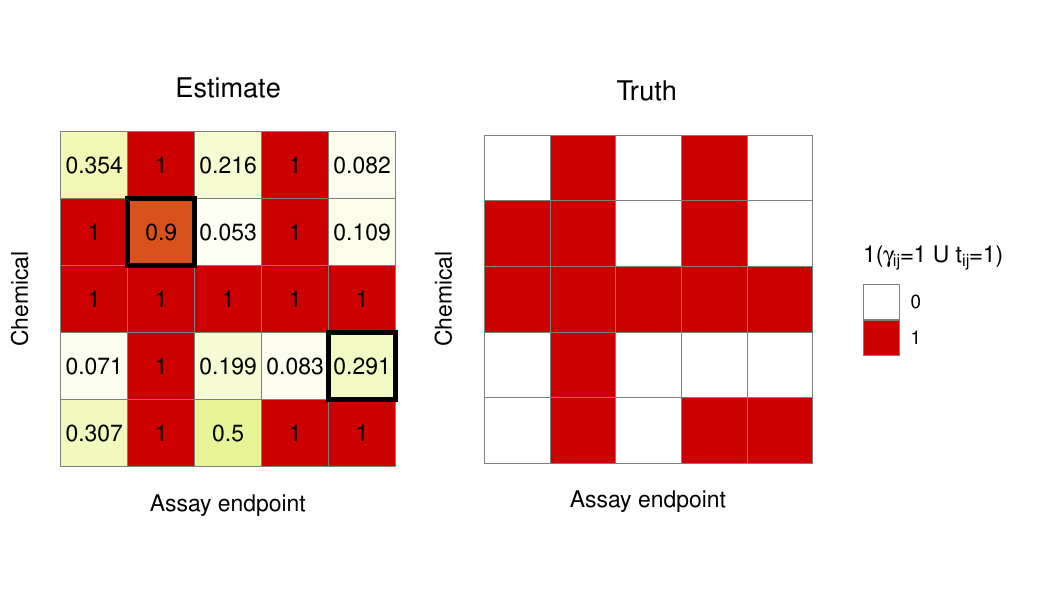}
\caption{Heat map of estimated and true profiles of activity from Simulation 1. Figure presents the results from a $5 \times 5$ subset chosen at random. The value in each cell of the left panel is the posterior mean of $\bm{1}(\gamma_{ij}=1\cup t_{ij}=1)$. Cells with outer lines ((2,2) and (4,5) elements) are held-out pairs for which $\bm{1}(\gamma_{ij}=1\cup t_{ij}=1)$'s are predicted.}
\label{Sim1:gamprofile}
\end{figure}

BMC provides precise estimation of the latent correlations among chemicals (Figure \ref{Sim1:corr} in the Supporting Material). Two factors ($q=2$) generated the truth, and the sampler ran with a guess of three more factors. The multiplicative gamma process shrinkage prior helped recover the true number of factors $q=2$ by shrinking factor loadings of redundant factors to zero (Figure \ref{Sim1:loadings} in the Supporting Material). Figure \ref{Sim1:gamprofile} displays an example of activity profiles. The truth is adequately captured via the estimated and predicted probabilities. Results from a $5 \times 5$ subset of the whole heat map are shown for better visualisation. The complete matrices of estimates and the truth are quite similar. 

Table \ref{table:simulation1} summarises simulation results when the data generating process is BMC. From BMC-variants, results from BMC$_i$ are presented because BMC$_i$ showed slightly better performance over the other two. Note that Area Under the ROC Curves (AUCs) from tcpl in Table \ref{table:simulation1} \& \ref{Simulation2} were computed slightly differently than those from other methods. BMC, three variations, and ZIPLL all produce \textit{probability} of active responses, which can be any value between 0 and 1. In order to evaluate the accuracy of estimates compared to the true $\gamma_{ij} \in \{0,1\}$ values, ROC curves and the corresponding AUCs are computed by changing thresholds between 0 and 1. On the other hand, EPA provides a \textit{binary} hit-call variable for the mean effect through ToxCast/Tox21. We hereafter refer to this variable (based on the version invitroDBv2) as EPA's hit-call. The EPA's hit-call identifies a pair as active if the fitted Hill or gain-loss model have lower Akaike information criterion than a constant model, and both the estimated and observed maximum responses exceed an efficacy cutoff chosen for the assay endpoint. This classification of whether each pair is active or not is directly comparable to the true $\gamma_{ij}$ without changing thresholds. In simulations, assay endpoint-specific cutoffs are set at 0.

Table \ref{table:simulation1} shows that BMC outperforms the other methods overall. In training data sets, BMC approaches (BMC and BMC$_i$) have lower RMSEs and higher AUCs compared to tcpl or ZIPLL. Poor performance of ZIPLL in these simulations is partially due to the facts that monotone increasing shape restrictions fail to fit decreasing trends and that ZIPLL does not allow for different $\sigma^2_j$'s. BMC outperforming tcpl may be due to the borrowing of information across chemicals and assay endpoints. Another benefit of BMC is the capability of modelling heteroscedasticity. The AUCs for $t_{ij}$ in Table \ref{table:simulation1} exhibit highly accurate estimation/prediction performance for detecting potential heteroscedastic variances, which is not available for non-BMC models. Moreover, BMC produces in- and out-of-sample AUCs that are uniformly better than those from BMC$_i$. Hence, when the factor model provides a realistic characterisation of the dependence structure across assay endpoints and chemicals, it is not suggested to use a simplified model for multiple testing. Less structure in $\gamma_{ij}$ and $t_{ij}$ results in lower out-of-sample AUCs. 

\begin{table}
\caption{Mean and standard errors in parenthesis across 30 simulation results when BMC is the true data generating process. Root Mean Squared Error (RMSE), and Area Under the ROC curve (AUC) results for probabilities of mean effect, variance effect, and activity are presented.}
\begin{tabular}{c|ccccc}
     & BMC & BMC$_i$ & ZIPLL & tcpl  \\ \hline
    RMSE & 0.423 (0.016) & 0.424 (0.016) & 0.834 (0.028) & 0.696 (0.015)\\
    In-sample AUC for $\gamma_{ij}$ & 0.995 (0.001) & 0.991 (0.002) & 0.667 (0.006) & 0.811 (0.008) \\
    Out-of-sample AUC for $\gamma_{ij}$ & 0.786 (0.068) & 0.502 (0.043) & - & - \\
    In-sample AUC for $t_{ij}$ & 0.999 (0.001) & 0.998 (0.001) & - & -\\
    Out-of-sample AUC for $t_{ij}$ & 0.794 (0.068) & 0.503 (0.029) & -& - \\
    In-sample AUC for& \multirow{2}{*}{$>$0.999 ($<$0.001)
} & \multirow{2}{*}{0.999 ($<$0.001)} & \multirow{2}{*}{-} & \multirow{2}{*}{-}\\
    $\bm{1}(\gamma_{ij} = 1 \cup t_{ij} = 1)$ &&&&\\
    Out-of-sample AUC for& \multirow{2}{*}{0.828 (0.064)} & \multirow{2}{*}{0.523 (0.053)} & \multirow{2}{*}{-} &\multirow{2}{*}{-} \\
    $\bm{1}({\gamma_{ij} = 1 \cup t_{ij} = 1})$ & &  &  &
\end{tabular}
\label{table:simulation1}
\end{table}

Figure \ref{Sim1:curves} illustrates that BMC closely recovers the true curves even in the existence of heteroscedasticity (\textsf{A} and \textsf{B}). To see if there are any unexplained patterns in residuals, plots of the residuals versus fitted values were examined from \textsf{A} and \textsf{B} (See Figure \ref{Sim1:fitvsres} in the Supporting Material). The ZIPLL does not consider heteroscedasticity in the model and consequently results in heteroscedastic residuals. In contrast, BMC is able to properly account for heteroscedasticity, and residuals do not show any patterns against fitted values. In addition, BMC nicely differentiates variance changes and mean changes. For instance, the estimated probability of the variance effect is around 0.3, while the probability of the mean effect is 1 in \textsf{C}.

Simulation 2 generates data from an alternative model, ZIPLL. Despite misalignment in data structure assumed by BMC and by ZIPLL, BMC performs similarly to ZIPLL and outperforms tcpl with respect to RMSE and AUC. The high in-sample AUC for $\gamma_{ij}$ from BMC (0.982) suggests its stable estimation performance even with relatively small number of chemicals and assay endpoints ($m=J=15$) and model misspecification. We provide a full discussion of Simulation 2 results in the Supporting Material \ref{s:sims}.
    
Another simulation (Simulation 3) is conducted to show how and where multiplicity adjustment occurs. We ran simulations with the number of chemicals $m=5$ and assay endpoints $J=5$ and repeat for increasing $J=6, 20, 50, 100$. Data were generated assuming BMC is the true model. Throughout the simulations, the number of 1's in $\gamma_{ij}$ and in $t_{ij}$ are fixed at 20 and 18, respectively. We expect our testing framework to have a strong control over false positives. Table \ref{tab:multiplicity} shows that the false positives evaluated at 0.5 remain small and steady as $J$ increases. We also observe decreasing posterior means of $\alpha_0$ and $\xi$ towards large negative values as expected. Therefore, we conclude that the multiple testing problem is properly accounted for under BMC. More simulation settings for Simulation 3 are fully described in the Supporting Material \ref{s:sims}. 

\begin{table}
\caption{Summary of results for multiplicity simulations with $m=5$ fixed and $J$ increasing. False positives (FPs) are computed at the cutoff 0.5 (if the posterior probability exceeds 0.5 then it is taken as a ``positive''). Posterior mean of $\alpha_0$ and $\xi$ are provided.}
\centering
    \begin{tabular}{c|cccc}
            & FPs for $\gamma_{ij}$ & FPs for $t_{ij}$ & $\alpha_0$ & $\xi$   \\ \hline
    $J=5$   & 3                    & 0               & 0.209      & 1.824 \\
    $J=6$   & 2                    & 2               & 0.537      & 0.959 \\
    $J=20$  & 6                    & 1               & -1.922     & -0.463 \\
    $J=50$  & 3                    & 2               & -2.481     & -1.032 \\
    $J=100$ & 2                    & 3               & -3.446     & -1.324
    \end{tabular}
    \label{tab:multiplicity}
\end{table}

In a final set of simulations, we empirically investigate performance of $Pr(\gamma_{ij} = 1 ~\cup~ t_{ij} = 1)$ in terms of AUC by varying missingness and correlation structures (Simulation 4). We examine two cases in which $m=30$ chemicals are highly correlated or weakly correlated (See Figure \ref{fig:sim7_corr} in the Supporting Material). Missingness varies from 10\% to 50\% in which the maximum level is chosen to reflect ToxCast/Tox21 data. As expected, out-of-sample AUCs tend to decrease as missingness increases. Despite the decreasing trend, the out-of-sample AUCs remain high ($>0.9$) even with 50\% missingness when chemicals are highly correlated and the model is well-specified. This is due to the fact that BMC properly exploits the correlated structure of chemicals through a latent factor model. However, performance can decline in weak correlation cases as borrowing of information pays less dividends; out-of-sample AUCs are acceptable ($>0.7$) only up to 30\% missingness. These results are summarised in Table \ref{tab:missing1} and \ref{tab:missing2} in the Supporting Material. 
    
\subsection{ToxCast/Tox21 Results} \label{ToxCast}
This section presents results from the ToxCast/Tox21 data analysis with a focus on endpoints relevant to human neurodevelopmental disorders and obesity. We ran the sampler for 40,000 iterations from which 30,000 were discarded as burn-in, and every 10th sample was saved for the next 10,000 iterations. This long burn-in is to be conservative; trace plots and effective sample sizes for posterior samples indicated good mixing and apparent convergence after 15,000 iterations. We also checked the log-linear assumption for heteroscedastic variances using posterior predictive samples. We computed the empirical coverage of 95\% posterior predictive intervals for data points in cells with estimated $t_{ij} > 0.5$. The average is 0.961, and the middle 50\% of the empirical coverage rates lie in [0.941, 0.989]. Therefore, we conclude that the log-linear variance model provides an adequate fit to the data.

We provide a new mean activity indicator $\kappa_{ij}$ along with $\gamma_{ij}$ for the ToxCast/Tox21 analysis. The $\kappa_{ij}$ is designed to incorporate efficacy cutoffs motivated by the EPA's hit-call. Efficacy cutoffs are specific to assay endpoints and provide a minimum magnitude for biologically interesting maximal responses. Recall that the EPA's hit-call is 1 if the Hill or gain-loss model wins over a constant model, and both the estimated and observed maximum responses are larger than an efficacy cutoff for each assay endpoint. The rationale behind the two criteria for EPA's hit-calls is to incorporate scientific significance through efficacy cutoffs as well as statistical significance. In the ToxCast/Tox21 data, $\xi$ is estimated as $1.147$ with the 95\% CI $(1.046, 1.247)$, yielding $\Phi(\hat{\xi}) = 0.874$. This high intercept suggests that many small signals are statistically significant with naive $\gamma_{ij}$. Hence, we make $\kappa_{ij} = \gamma_{ij}\bm{1}(\max{(X_{ij}\bm{\beta}_{ij})} > \text{cutoff}_{ij})$, that is, $\kappa_{ij}$ is 1 if $\gamma_{ij}=1$ \textit{and} the fitted maximum exceeds a cutoff. This is designed to utilise scientific knowledge about \textit{big enough} signals. In our analyses, we normalise data within each pair, so the cutoffs are also normalised accordingly. We expect $\kappa_{ij}$ to be more conservative than $\gamma_{ij}$. BMC provides both metrics $\kappa_{ij}$ and $\gamma_{ij}$ so that researchers can have balanced understanding of chemicals' mean activity based on scientific and statistical significance. 

Despite $\gamma_{ij}$ being sensitive to small signals, $t_{ij}$ is much more robust and conservative even though they share the same factor loadings and latent factors. The difference in behaviours is attributable to $\bm{\alpha}$ coefficients. In the ToxCast/Tox21 application, $\alpha_0$ is $-0.208$ with 95\% CI $(-0.271,-0.140)$, and $\alpha_1$ is negative with mean $-2.610$ with 95\% CI $(-3.138, -2.171)$. This negative value may not imply that mean activity and variance activity move in different directions. Rather, we view it as undoing detection of cells with small signals as they are unlikely to be disturbed in variance. This perspective is supported by that converted 0/1's from posterior means of $\gamma_{ij}$ and $t_{ij}$ achieve strong concordance when higher conversion cutoff is applied to $\gamma_{ij}$ than to $t_{ij}$.

We discovered that one of the latent factors of assay endpoints is highly related to detection technology (See Figure \ref{obese:latfac}). Measuring assay endpoints relies on a variety of different technologies for detecting and quantifying analytes; these technologies have different levels of sensitivity to bioactivity. Therefore, it is reasonable that one of the activity-relevant latent factors has a strong connection with detection technology.

\begin{figure}[htbp]
\centering
\includegraphics[height=0.9\textheight]{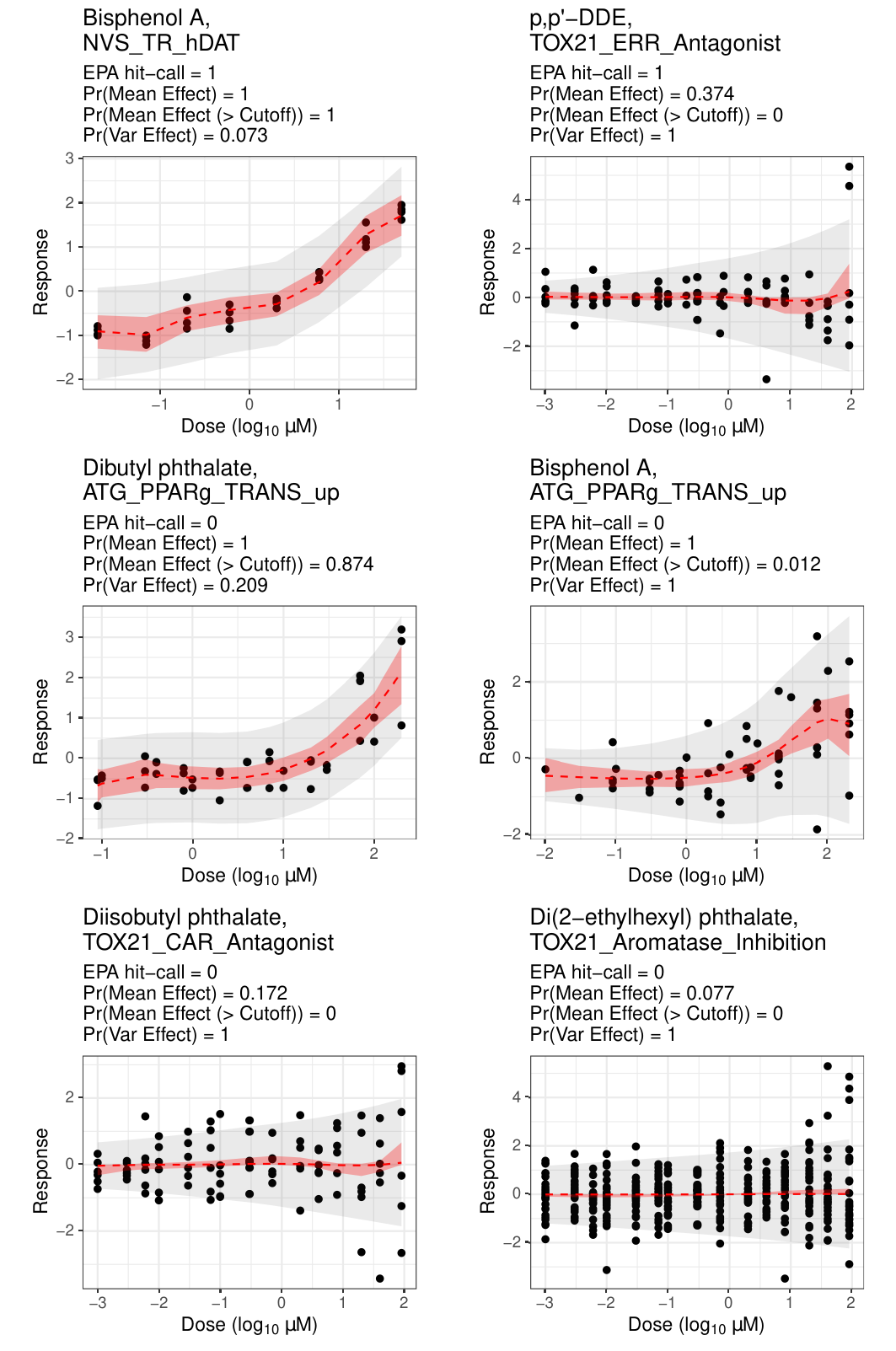}
\caption{Fitted results for select chemical-assay endpoint pairs estimated to be active by BMC.}
\label{hitcbad}
\end{figure}

Figure \ref{hitcbad} shows estimated dose-response curves from BMC as dashed lines with 95\% CIs as shaded areas in red. The light gray shaded areas illustrate 95\% posterior predictive intervals for the data points drawn as black dots. ``Pr(Mean Effect)'' is the mean effect probability for a (chemical $i$, assay endpoint $j$) pair, which is computed as the posterior mean of $\gamma_{ij}$. ``Pr(Mean Effect ($>$ Cutoff))'' is a more conservative measure, the probability of mean effects exceeding a cutoff, which is the posterior mean of $\kappa_{ij}$. Similarly, ``Pr(Var Effect)'' indicates the variance effect probability whose value is the posterior mean of $t_{ij}.$ 

The first row of Figure \ref{hitcbad} shows that BMC is able to differentiate dose effects on the mean from dose effects on the variance of dose-response curves. Recall that the EPA's hit-call is an indication of mean changes. In the left panel, BMC and the EPA agree that mean changes exist, which is supported by an increasing trend. In the right panel, the EPA's hit-call claims that the average dose-response is not constant. However, BMC estimates that the mean curve is likely to be constant at zero, but with there being clear evidence of heteroscedasticity. Therefore, the first row in Figure \ref{hitcbad} suggests that (1) BMC can separate mean and variance effects (at least in some cases); and (2) the EPA's hit-call might be misled by heteroscedastic variances.

The second row of Figure \ref{hitcbad} illustrates some cases where BMC and the EPA's hit-call disagree, and BMC's result is more plausible. For both pairs, the EPA's hit-calls say no activity because their fitted maximum responses by Hill model do not exceed the assay endpoint's efficacy cutoff (1.174). However, BMC estimates both pairs to be active with high probability. Notice that posterior summaries of $\gamma_{ij}$ and $\kappa_{ij}$ agree on the left panel, while $\kappa_{ij}$ drastically drops compared to $\gamma_{ij}$ on the right. Nonetheless, it is evident that BPA induces heteroscedastic responses. In fact, for these two chemicals, Dibutyl phthalate and BPA, not only do plots show perturbations in dose-response measurements, but also background knowledge supports BMC's estimates. First, BPA and phthalates are known to disrupt the endocrine system, which potentially results in neurodevelopmental disorders \citep{tran2017neurodevelopmental} and obesity \citep{holtcamp2012obesogens}. Second, chemicals activating PPAR$_{\gamma}$ (or PPARg) receptors are potential obesogens because PPAR$_{\gamma}$ is a master regulator in formulating fat cells \citep{evans2004ppars}. To be specific, ATG\_PPAR$_{\gamma}$\_TRANS\_up represents an endpoint captured through a human liver cell-based assay. The mechanism of action for obesogens like BPA altering PPAR$_{\gamma}$ activity in the liver is well-established \citep{marmugi2012low, diamante2021systems}. Therefore, it is not unexpected for BPA and Dibutyl phthalate to be active for the given assay endpoint, ATG\_PPAR$_{\gamma}$\_TRANS\_up. 

The third row of Figure \ref{hitcbad} shows cases where the EPA's hit-call can have low power because it misses signals manifest in the variance instead of the mean. Given that phthalates are related to obesity \citep{holtcamp2012obesogens}, we expect disruptive patterns on assay endpoints presenting toxicity of Diisobutyl phthatlate and Di(2-ethylhexyl) phthatlate. However, the EPA's hit-call suggests that these phthalates are not active at the doses tested. This may be true in terms of mean changes, but variances seem clearly heteroscedastic. 

\begin{figure}[htbp]
\centering
\includegraphics[width=\textwidth]{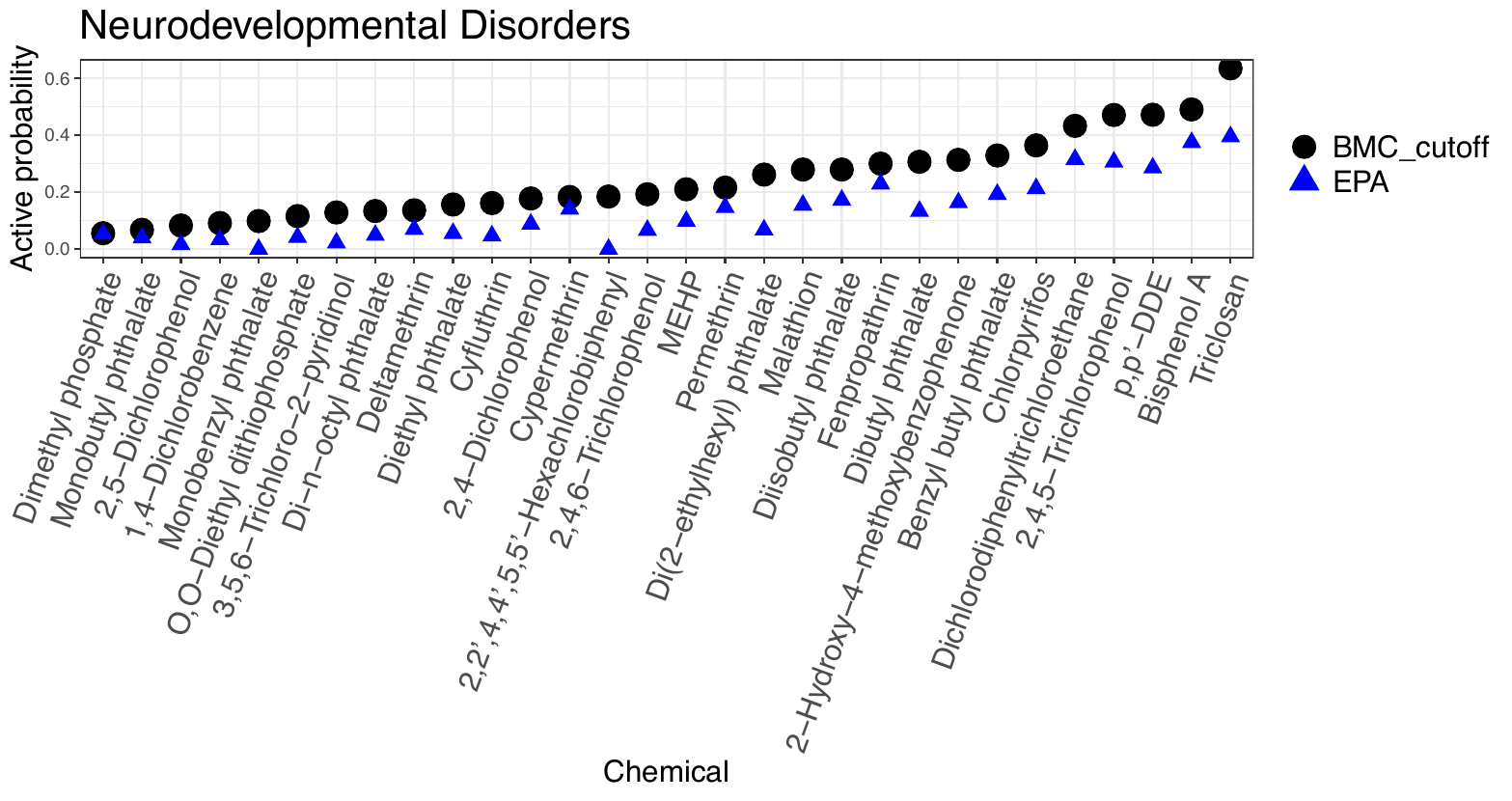}
\caption{Chemical ranks by the average active probability from BMC (dots) and the average hit-call from EPA (triangles) over assay endpoints related to neurodevelopmental disorders.}
\label{neuro_disruptchem}
\end{figure}

\begin{figure}[htbp]
\centering
\includegraphics[width=\textwidth]{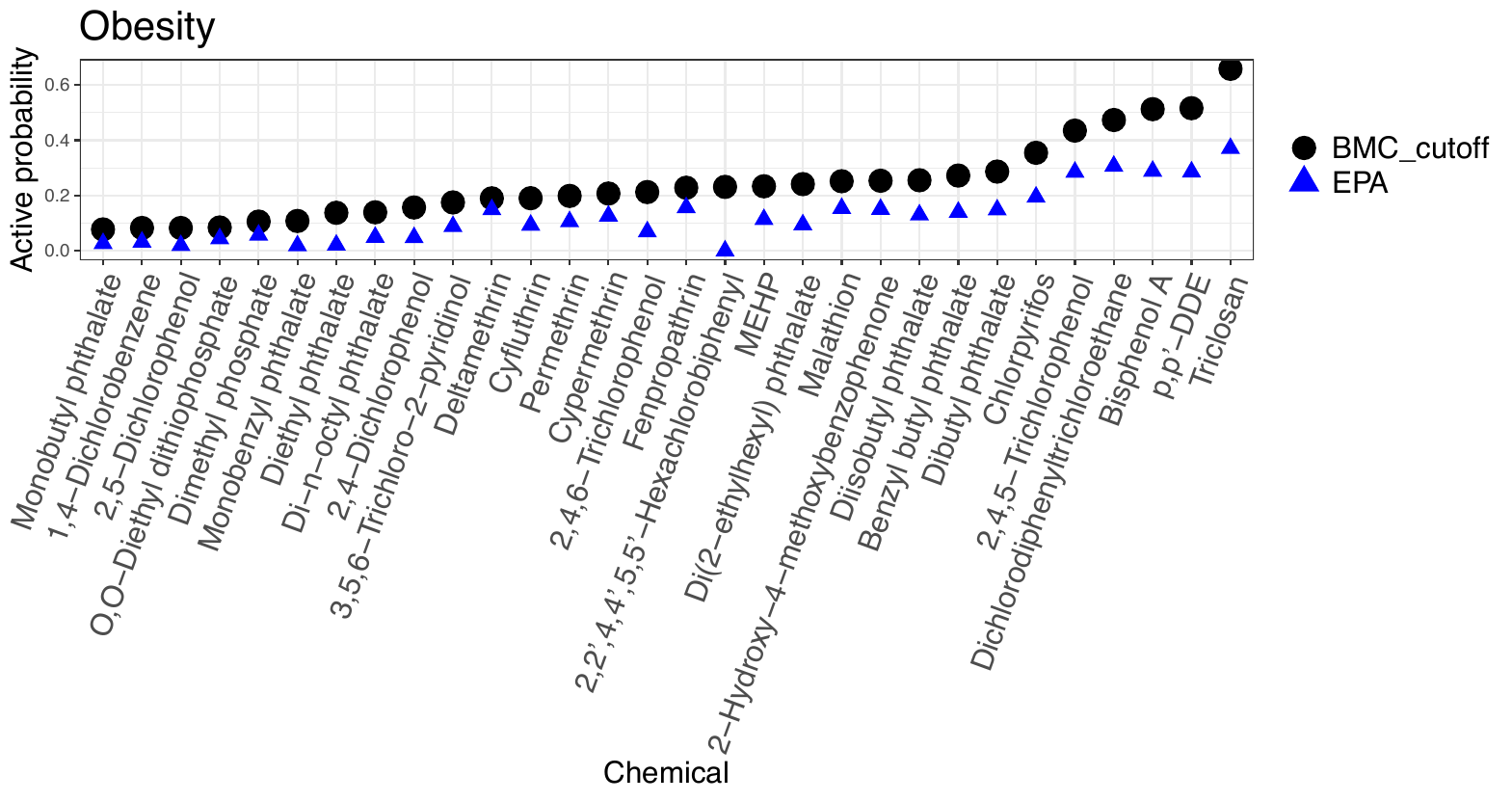}
\caption{Chemical ranks by the average activity probability from BMC (dots) and the average hit-call from EPA (triangles) over obesity-related assay endpoints.}
\label{obese_disruptchem}
\end{figure}

Figure \ref{neuro_disruptchem} and \ref{obese_disruptchem} show chemicals in order of average activity probability with cutoffs, $Pr(\kappa_{ij}=1\cup t_{ij}=1)$, over assay endpoints related to neurodevelopmental disorders and obesity, respectively. Top five chemicals that are most likely to disrupt biological processes associated with the two diseases are Triclosan, p,p'-DDE, BPA, Dichlorodiphenyltrichloroethane (DDT), and 2,4,5-Trichlorophenol. In the figures, we notice that the rankings of chemicals by BMC and by the EPA's hit-call show only subtle differences. But, the actual probabilities computed by BMC are uniformly higher than average hit-calls from EPA. This is because (1) BMC's active probabilities include heteroscedasticity, while EPA's hit-call detects the mean activity only; and (2) BMC improves power of tests by borrowing information across multiple chemicals and assay endpoints. Since these bioactivity rankings are based on the data that are currently available, it will be informative to revisit such rankings as data expand.

To study sensitivity of rankings to the choice of chemicals, we expanded our analysis to 326 chemicals. They consist of the original 30 chemicals and those screened in Phase I of the ToxCast that have been exclusively used in other toxicity studies including \cite{martin2010impact} and \cite{wilson2014hierarchical}. Within this larger collection, relative positions of the 30 chemicals remained intact with a few exceptions. BPA and Triclosan were positioned lower in the larger set, while Cyfluthrin and MEHP were positioned higher. One of explanations for these shifts is an altered correlation structure among chemicals. The Phase I chemicals are mostly pesticides, and the four chemicals might have different relationships with pesticides from what they had with the 30 chemicals in terms of the mean effect. 

One can construct a list of assay endpoints highly likely to be activated by the most active chemicals regarding disease outcomes of interest. Then the assay endpoints in the list are expected to have important implications in progression for the diseases. The ToxCast/Tox21 data include both agonist and antagonist assays, and thus the activated probability encompasses agonist and antagonist directions. Such lists for neurodevelopmental disorders and obesity are available in Figure \ref{neuro_disruptassay} and \ref{obese_disruptassay} in the Supporting Material, from which fifteen assay endpoints show impacts on both disease classes. 

\section{Discussion} \label{Discussion}
    
We have proposed a Bayesian multiple testing approach for inference on activity of chemicals in settings involving multiple chemicals and assay endpoints and possible heteroscedasticity. Our BMC approach can be applied directly in other settings involving a similar matrix-structured experimental design. For example, this is common in pharmaceutical studies assessing drug activity, which will look for evidence of activity for different health outcomes. Also, in microbial genetics, similar designs are conducted but for different types of bacteria and environmental conditions. 

The ultimate goal of many analyses using \textit{in vitro} data is to make inferences on human health and inform protective regulations. Accordingly, chemicals and assay endpoints studied in the ToxCast/Tox21 application are carefully selected: the chemicals are also measured in human epidemiology studies, and the assay endpoints cover a variety of species and several types of tissue targets. Especially, we identified assay endpoints for various pathways relevant to specific disease outcomes across multiple data sources. This identification is a highly valuable and meaningful practice in various fields of science including epidemiology and toxicology in that it is adaptive to other disease outcomes and extendable to different data sources as well. 

Using the \textit{in vitro} data, we were able to rank chemicals based on their active probabilities and find the most disruptive chemicals including, but not limited to, Triclosan, p,p'-DDE, BPA, DDT, and 2,4,5-Trichlorophenol. It will be interesting to follow up on these top ranking chemicals for neurodevelopmental disorders and obesity disease outcomes to further elucidate their role in human health. In the future, we may consider including chemicals' molecular structure information to increase power of hypothesis tests, motivated by several  Quantitative Structure–Activity Relationship (QSAR) models \citep{low2015bayesian, wheeler2019bayesian, moran2019bayesian}. 

When extending \textit{in vitro} results to \textit{in vivo} toxicity, doses need to be carefully considered. All the results presented in the paper should be interpreted in terms of tested doses, so we do not conclude a chemical with a high probability of inactivity is inactive at higher doses than those tested. Simultaneously, it is recommended to ensure that the doses tested \textit{in vitro} can physiologically occur in animals/humans. This recommendation is reinforced by \cite{klaren2019identifying} in which \textit{in vivo} toxicity prediction using \textit{in vitro} assays performs much better with toxicokinetic modelling. Therefore, future research linking \textit{in vitro} data and \textit{in vivo} implications could be greatly assisted by assuring dose applicability in animals/humans as well as widening the range of tested doses. 

\section*{Acknowledgments}
We are grateful for the financial support of the National Institute of Environmental Health Sciences through grants R01ES027498 and R01ES028804. The authors would like to thank Brett Winters for identifying assay endpoints relevant to neurodevelopmental disorders and obesity, Evan Poworoznek for sharing computer code, Kelly Moran for helpful comments, and Matthew Wheeler for help in processing ToxCast/Tox21 data. We deeply appreciate helpful comments from the editor, the associate editor and two referees. We thank the first referee for the suggestion of using a factor model for heteroscedasticity.
\vspace*{-8pt}

\bibliographystyle{rss}
\bibliography{bibliography}

\begin{thebibliography}{40}
\expandafter\ifx\csname natexlab\endcsname\relax\def\natexlab#1{#1}\fi
\expandafter\ifx\csname url\endcsname\relax
  \def\url#1{\texttt{#1}}\fi
\expandafter\ifx\csname urlprefix\endcsname\relax\def\urlprefix{URL: }\fi

\bibitem[{Bhattacharya and Dunson(2011)}]{bhattacharya2011sparse}
Bhattacharya, A. and Dunson, D.~B. (2011) Sparse {Bayesian} infinite factor
  models.
\newblock \textit{Biometrika}, \textbf{98}, 291--306.

\bibitem[{Corty and Valdar(2018)}]{corty2018vqtl}
Corty, R.~W. and Valdar, W. (2018) Vqtl: an {R} package for mean-variance {QTL}
  mapping.
\newblock \textit{G3: Genes, Genomes, Genetics}, \textbf{8}, 3757--3766.

\bibitem[{Davis et~al.(2019)Davis, Grondin, Johnson, Sciaky, McMorran, Wiegers,
  Wiegers and Mattingly}]{davis2019comparative}
Davis, A.~P., Grondin, C.~J., Johnson, R.~J., Sciaky, D., McMorran, R.,
  Wiegers, J., Wiegers, T.~C. and Mattingly, C.~J. (2019) The comparative
  toxicogenomics database: update 2019.
\newblock \textit{Nucleic {A}cids {R}esearch}, \textbf{47}, D948--D954.

\bibitem[{Diamante et~al.(2021)Diamante, Cely, Zamora, Ding, Blencowe, Lang,
  Bline, Singh, Lusis and Yang}]{diamante2021systems}
Diamante, G., Cely, I., Zamora, Z., Ding, J., Blencowe, M., Lang, J., Bline,
  A., Singh, M., Lusis, A.~J. and Yang, X. (2021) Systems toxicogenomics of
  prenatal low-dose {BPA} exposure on liver metabolic pathways, gut microbiota,
  and metabolic health in mice.
\newblock \textit{Environment International}, \textbf{146}, 106260.

\bibitem[{Dix et~al.(2007)Dix, Houck, Martin, Richard, Setzer and
  Kavlock}]{dix2007ToxCast}
Dix, D.~J., Houck, K.~A., Martin, M.~T., Richard, A.~M., Setzer, R.~W. and
  Kavlock, R.~J. (2007) The {ToxCast} program for prioritizing toxicity testing
  of environmental chemicals.
\newblock \textit{Toxicological Sciences}, \textbf{95}, 5--12.

\bibitem[{Durante(2017)}]{durante2017note}
Durante, D. (2017) A note on the multiplicative gamma process.
\newblock \textit{Statistics \& Probability Letters}, \textbf{122}, 198--204.

\bibitem[{ECHA(2017)}]{echa2017use}
ECHA (2017) The use of alternatives to testing on animals for the {REACH}
  regulation.
\newblock \textit{European Chemicals Agency, Helsinki, Finland},
  \url{https://doi.org/10.2823/023078}.

\bibitem[{Evans et~al.(2004)Evans, Barish and Wang}]{evans2004ppars}
Evans, R.~M., Barish, G.~D. and Wang, Y.-X. (2004) {PPARs} and the complex
  journey to obesity.
\newblock \textit{Nature Medicine}, \textbf{10}, 355--361.

\bibitem[{Filer et~al.(2017)Filer, Kothiya, Setzer, Judson and
  Martin}]{filer2017tcpl}
Filer, D.~L., Kothiya, P., Setzer, R.~W., Judson, R.~S. and Martin, M.~T.
  (2017) Tcpl: the {ToxCast} pipeline for high-throughput screening data.
\newblock \textit{Bioinformatics}, \textbf{33}, 618--620.

\bibitem[{Holtcamp(2012)}]{holtcamp2012obesogens}
Holtcamp, W. (2012) Obesogens: an environmental link to obesity.
\newblock \textit{Environmental Health Perspectives}, \textbf{120}, a62--a68.

\bibitem[{Hsieh et~al.(2015)Hsieh, Sedykh, Huang, Xia and Tice}]{hsieh2015data}
Hsieh, J.-H., Sedykh, A., Huang, R., Xia, M. and Tice, R.~R. (2015) A data
  analysis pipeline accounting for artifacts in {Tox21} quantitative
  high-throughput screening assays.
\newblock \textit{Journal of Biomolecular Screening}, \textbf{20}, 887--897.

\bibitem[{Huang et~al.(2014)Huang, Sakamuru, Martin, Reif, Judson, Houck,
  Casey, Hsieh, Shockley, Ceger et~al.}]{huang2014profiling}
Huang, R., Sakamuru, S., Martin, M.~T., Reif, D.~M., Judson, R.~S., Houck,
  K.~A., Casey, W., Hsieh, J.-H., Shockley, K.~R., Ceger, P. et~al. (2014)
  Profiling of the {Tox21} 10k compound library for agonists and antagonists of
  the estrogen receptor alpha signaling pathway.
\newblock \textit{Scientific Reports}, \textbf{4}, 1--9.

\bibitem[{Judson et~al.(2016)Judson, Houck, Martin, Richard, Knudsen, Shah,
  Little, Wambaugh, Woodrow~Setzer, Kothya et~al.}]{judson2016editor}
Judson, R., Houck, K., Martin, M., Richard, A.~M., Knudsen, T.~B., Shah, I.,
  Little, S., Wambaugh, J., Woodrow~Setzer, R., Kothya, P. et~al. (2016)
  Editor's highlight: analysis of the effects of cell stress and cytotoxicity
  on in vitro assay activity across a diverse chemical and assay space.
\newblock \textit{Toxicological Sciences}, \textbf{152}, 323--339.

\bibitem[{Judson et~al.(2010)Judson, Houck, Kavlock, Knudsen, Martin,
  Mortensen, Reif, Rotroff, Shah, Richard et~al.}]{judson2010vitro}
Judson, R.~S., Houck, K.~A., Kavlock, R.~J., Knudsen, T.~B., Martin, M.~T.,
  Mortensen, H.~M., Reif, D.~M., Rotroff, D.~M., Shah, I., Richard, A.~M.
  et~al. (2010) In vitro screening of environmental chemicals for targeted
  testing prioritization: the {ToxCast} project.
\newblock \textit{Environmental {H}ealth {P}erspectives}, \textbf{118},
  485--492.

\bibitem[{Klaren et~al.(2019)Klaren, Ring, Harris, Thompson, Borghoff, Sipes,
  Hsieh, Auerbach and Rager}]{klaren2019identifying}
Klaren, W.~D., Ring, C., Harris, M.~A., Thompson, C.~M., Borghoff, S., Sipes,
  N.~S., Hsieh, J.-H., Auerbach, S.~S. and Rager, J.~E. (2019) Identifying
  attributes that influence in vitro-to-in vivo concordance by comparing in
  vitro {Tox21} bioactivity versus in vivo drugmatrix transcriptomic responses
  across 130 chemicals.
\newblock \textit{Toxicological Sciences}, \textbf{167}, 157--171.

\bibitem[{Knapen et~al.(2020)Knapen, Stinckens, Cavallin, Ankley, Holbech,
  Villeneuve and Vergauwen}]{knapen2020toward}
Knapen, D., Stinckens, E., Cavallin, J.~E., Ankley, G.~T., Holbech, H.,
  Villeneuve, D.~L. and Vergauwen, L. (2020) Toward an {AOP} network-based
  tiered testing strategy for the assessment of thyroid hormone disruption.
\newblock \textit{Environmental Science \& Technology}, \textbf{54},
  8491--8499.

\bibitem[{Koren et~al.(2009)Koren, Bell and Volinsky}]{koren2009matrix}
Koren, Y., Bell, R. and Volinsky, C. (2009) Matrix factorization techniques for
  recommender systems.
\newblock \textit{Computer}, \textbf{42}, 30--37.

\bibitem[{Leslie et~al.(2007)Leslie, Kohn and Nott}]{leslie2007general}
Leslie, D.~S., Kohn, R. and Nott, D.~J. (2007) A general approach to
  heteroscedastic linear regression.
\newblock \textit{Statistics and Computing}, \textbf{17}, 131--146.

\bibitem[{Li and Zhang(2010)}]{li2010bayesian}
Li, F. and Zhang, N.~R. (2010) Bayesian variable selection in structured
  high-dimensional covariate spaces with applications in genomics.
\newblock \textit{Journal of the American {S}tatistical {A}ssociation},
  \textbf{105}, 1202--1214.

\bibitem[{Low-Kam et~al.(2015)Low-Kam, Telesca, Ji, Zhang, Xia, Zink, Nel
  et~al.}]{low2015bayesian}
Low-Kam, C., Telesca, D., Ji, Z., Zhang, H., Xia, T., Zink, J.~I., Nel, A.~E.
  et~al. (2015) A {Bayesian} regression tree approach to identify the effect of
  nanoparticles’ properties on toxicity profiles.
\newblock \textit{Annals of Applied Statistics}, \textbf{9}, 383--401.

\bibitem[{Marmugi et~al.(2012)Marmugi, Ducheix, Lasserre, Polizzi, Paris,
  Priymenko, Bertrand-Michel, Pineau, Guillou, Martin et~al.}]{marmugi2012low}
Marmugi, A., Ducheix, S., Lasserre, F., Polizzi, A., Paris, A., Priymenko, N.,
  Bertrand-Michel, J., Pineau, T., Guillou, H., Martin, P.~G. et~al. (2012) Low
  doses of {Bisphenol A} induce gene expression related to lipid synthesis and
  trigger triglyceride accumulation in adult mouse liver.
\newblock \textit{Hepatology}, \textbf{55}, 395--407.

\bibitem[{Martin et~al.(2010)Martin, Dix, Judson, Kavlock, Reif, Richard,
  Rotroff, Romanov, Medvedev, Poltoratskaya et~al.}]{martin2010impact}
Martin, M.~T., Dix, D.~J., Judson, R.~S., Kavlock, R.~J., Reif, D.~M., Richard,
  A.~M., Rotroff, D.~M., Romanov, S., Medvedev, A., Poltoratskaya, N. et~al.
  (2010) Impact of environmental chemicals on key transcription regulators and
  correlation to toxicity end points within {EPA}’s toxcast program.
\newblock \textit{Chemical Research in Toxicology}, \textbf{23}, 578--590.

\bibitem[{Mnih and Salakhutdinov(2008)}]{mnih2008probabilistic}
Mnih, A. and Salakhutdinov, R.~R. (2008) Probabilistic matrix factorization.
\newblock In \textit{Advances in Neural Information Processing Systems},
  1257--1264.

\bibitem[{Moran et~al.(forthcoming)Moran, Dunson and
  Herring}]{moran2019bayesian}
Moran, K.~R., Dunson, D.~B. and Herring, A.~H. (forthcoming) Bayesian joint
  modeling of chemical structure and dose response curves.
\newblock \textit{Annals of Applied Statistics}.

\bibitem[{Neelon and Dunson(2004)}]{neelon2004bayesian}
Neelon, B. and Dunson, D.~B. (2004) Bayesian isotonic regression and trend
  analysis.
\newblock \textit{Biometrics}, \textbf{60}, 398--406.

\bibitem[{Par{\'e} et~al.(2010)Par{\'e}, Cook, Ridker and
  Chasman}]{pare2010use}
Par{\'e}, G., Cook, N.~R., Ridker, P.~M. and Chasman, D.~I. (2010) On the use
  of variance per genotype as a tool to identify quantitative trait interaction
  effects: a report from the women's genome health study.
\newblock \textit{PLoS {G}enetics}, \textbf{6}.

\bibitem[{Polson et~al.(2013)Polson, Scott and Windle}]{polson2013bayesian}
Polson, N.~G., Scott, J.~G. and Windle, J. (2013) Bayesian inference for
  logistic models using {P}{\'o}lya--{G}amma latent variables.
\newblock \textit{Journal of the American statistical Association},
  \textbf{108}, 1339--1349.

\bibitem[{Purushotham et~al.(2012)Purushotham, Liu and
  Kuo}]{purushotham2012collaborative}
Purushotham, S., Liu, Y. and Kuo, C.-C.~J. (2012) Collaborative topic
  regression with social matrix factorization for recommendation systems.
\newblock In \textit{Proceedings of the 29th International Coference on
  International Conference on Machine Learning}, 691--698.

\bibitem[{Ritz(2010)}]{ritz2010toward}
Ritz, C. (2010) Toward a unified approach to dose--response modeling in
  ecotoxicology.
\newblock \textit{Environmental Toxicology and Chemistry}, \textbf{29},
  220--229.

\bibitem[{R{\"o}nneg{\aa}rd and Valdar(2012)}]{ronnegaard2012recent}
R{\"o}nneg{\aa}rd, L. and Valdar, W. (2012) Recent developments in statistical
  methods for detecting genetic loci affecting phenotypic variability.
\newblock \textit{BMC {G}enetics}, \textbf{13}, 63.

\bibitem[{Scheel et~al.(2013)Scheel, Ferkingstad, Frigessi, Haug, Hinnerichsen
  and Meze-Hausken}]{scheel2013bayesian}
Scheel, I., Ferkingstad, E., Frigessi, A., Haug, O., Hinnerichsen, M. and
  Meze-Hausken, E. (2013) A {Bayesian} hierarchical model with spatial variable
  selection: the effect of weather on insurance claims.
\newblock \textit{Journal of the Royal Statistical Society: Series C (Applied
  Statistics)}, \textbf{62}, 85--100.

\bibitem[{Scott and Berger(2006)}]{scott2006exploration}
Scott, J.~G. and Berger, J.~O. (2006) An exploration of aspects of {Bayesian}
  multiple testing.
\newblock \textit{Journal of {S}tatistical {P}lanning and {I}nference},
  \textbf{136}, 2144--2162.

\bibitem[{Scott and Berger(2010)}]{scott2010bayes}
--- (2010) Bayes and empirical-{Bayes} multiplicity adjustment in the
  variable-selection problem.
\newblock \textit{Annals of Statistics}, \textbf{38}, 2587--2619.

\bibitem[{Tansey et~al.(2019)Tansey, Tosh and Blei}]{tansey2019relational}
Tansey, W., Tosh, C. and Blei, D.~M. (2019) A {B}ayesian model of dose-response
  for cancer drug studies.
\newblock \textit{arXiv preprint}, \url{https://arxiv.org/abs/1906.04072}.

\bibitem[{Thomas et~al.(2009)Thomas, Conti, Baurley, Nijhout, Reed and
  Ulrich}]{thomas2009use}
Thomas, D.~C., Conti, D.~V., Baurley, J., Nijhout, F., Reed, M. and Ulrich,
  C.~M. (2009) Use of pathway information in molecular epidemiology.
\newblock \textit{Human {G}enomics}, \textbf{4}, 21.

\bibitem[{Tice et~al.(2013)Tice, Austin, Kavlock and
  Bucher}]{tice2013improving}
Tice, R.~R., Austin, C.~P., Kavlock, R.~J. and Bucher, J.~R. (2013) Improving
  the human hazard characterization of chemicals: a {Tox21} update.
\newblock \textit{Environmental Health Perspectives}, \textbf{121}, 756--765.

\bibitem[{Tran and Miyake(2017)}]{tran2017neurodevelopmental}
Tran, N. Q.~V. and Miyake, K. (2017) Neurodevelopmental disorders and
  environmental toxicants: Epigenetics as an underlying mechanism.
\newblock \textit{International Journal of Genomics}, \textbf{2017}, 1--23.

\bibitem[{Wheeler(2019)}]{wheeler2019bayesian}
Wheeler, M.~W. (2019) Bayesian additive adaptive basis tensor product models
  for modeling high dimensional surfaces: an application to high-throughput
  toxicity testing.
\newblock \textit{Biometrics}, \textbf{75}, 193--201.

\bibitem[{Wilson et~al.(2014)Wilson, Reif and Reich}]{wilson2014hierarchical}
Wilson, A., Reif, D.~M. and Reich, B.~J. (2014) Hierarchical dose--response
  modeling for high-throughput toxicity screening of environmental chemicals.
\newblock \textit{Biometrics}, \textbf{70}, 237--246.

\bibitem[{Yang et~al.(2012)Yang, Loos, Powell, Medland, Speliotes, Chasman,
  Rose, Thorleifsson, Steinthorsdottir, M{\"a}gi et~al.}]{yang2012fto}
Yang, J., Loos, R.~J., Powell, J.~E., Medland, S.~E., Speliotes, E.~K.,
  Chasman, D.~I., Rose, L.~M., Thorleifsson, G., Steinthorsdottir, V.,
  M{\"a}gi, R. et~al. (2012) {FTO} genotype is associated with phenotypic
  variability of body mass index.
\newblock \textit{Nature}, \textbf{490}, 267--272.

\end{thebibliography}

% Supplemental Material 
\clearpage
\begin{center}
\textbf{\large Web-based supporting materials for \\``Bayesian Matrix Completion for Hypothesis Testing''}
\end{center}
\setcounter{section}{0}
\setcounter{equation}{0}
\setcounter{figure}{0}
\setcounter{table}{0}
\setcounter{page}{1}
\makeatletter
\renewcommand{\thesection}{S\arabic{section}}
\renewcommand{\theequation}{S\arabic{equation}}
\renewcommand{\thefigure}{S\arabic{figure}}
\renewcommand{\thetable}{S\arabic{table}}
\renewcommand{\bibnumfmt}[1]{[S#1]}
\renewcommand{\citenumfont}[1]{S#1}

\section{Figures} \label{s:figure}
\begin{figure}[htbp]
\centering
\includegraphics[width=0.65\textwidth]{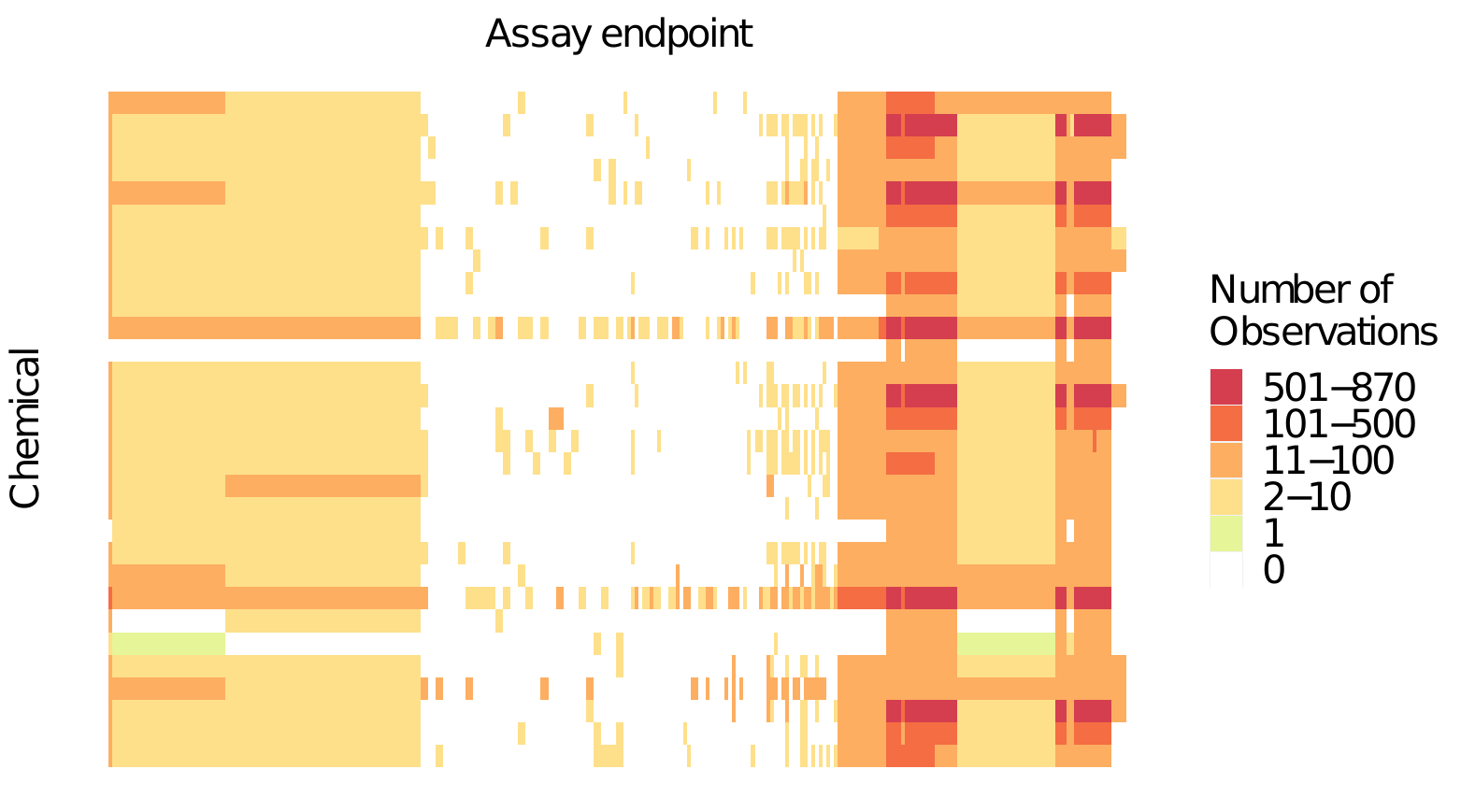}
\caption{Heat map of the number of observations in ToxCast/Tox21 data for obesity, based on 30 chemicals on rows and 271 assay endpoints on columns.}
\label{DataMatWhole}
\end{figure}

\begin{figure}
\centering
\includegraphics[width=0.65\textwidth]{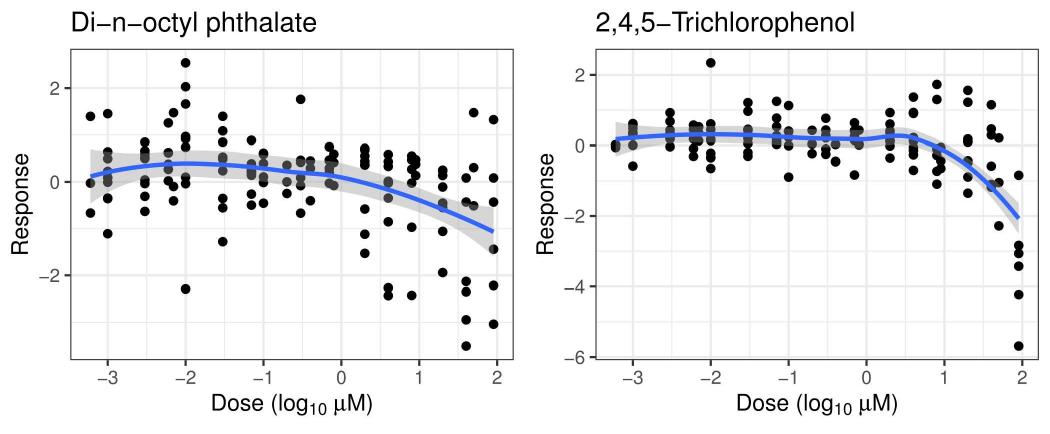}
\caption{Scatter plots of two chemicals on the TOX21\_ERa\_LUC\_BG1\_Agonist assay endpoint. The solid lines and gray shaded areas represent the average dose-response curves and 95\% confidence intervals fitted via locally estimated scatterplot smoothing (LOESS).}
\label{decreasingfunc}
\end{figure}

\begin{figure}[htbp]
\centering
\includegraphics[width=0.5\textwidth]{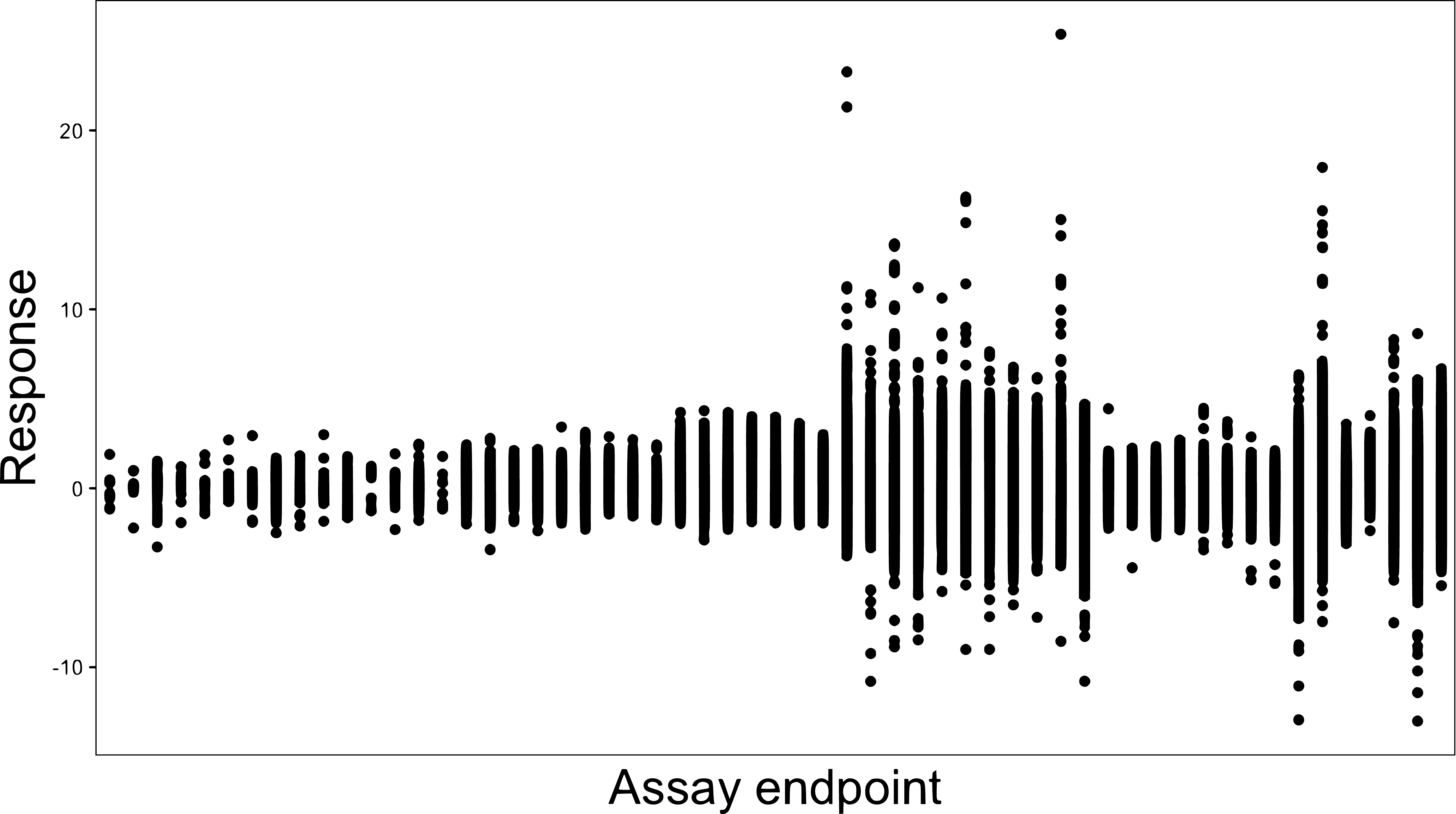}
\caption{Scatter plot of the responses normalised by chemical-assay endpoint pairs. Points on each vertical line represent responses from one assay endpoint. This figure is based on a subset of assay endpoints.}
\label{assayspecificvar}
\end{figure}

\begin{figure}[htbp]
\centering
\includegraphics[width=0.6\textwidth]{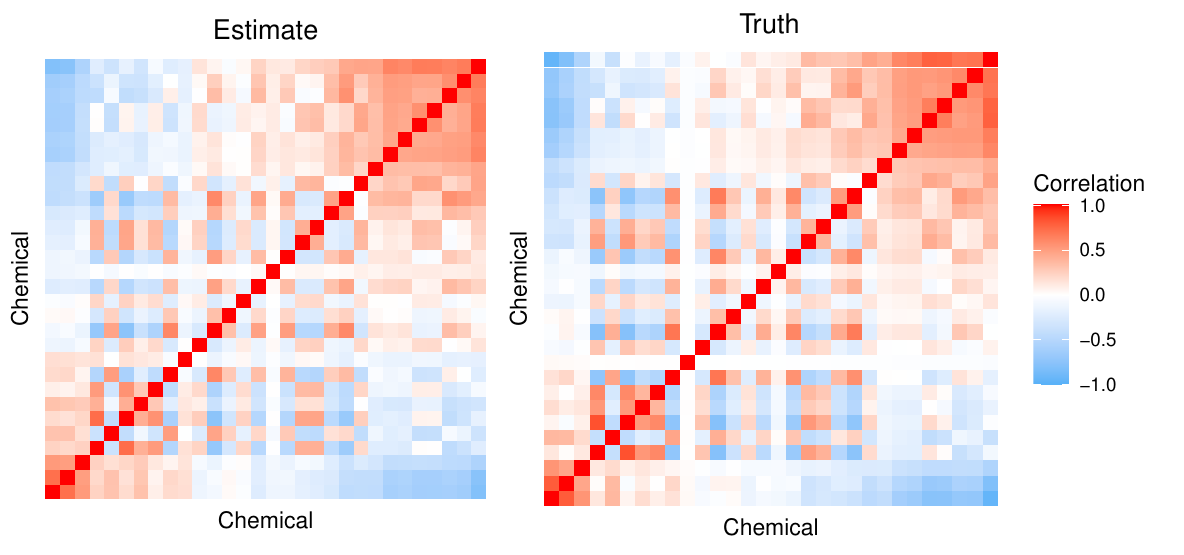}
\caption{Heat map of the estimated and true correlation matrix among chemicals with respect to the mean effect. The results are from Simulation 1.}
\label{Sim1:corr}
\end{figure}

\begin{figure}[htbp]
\centering
\includegraphics[height=0.35\textheight]{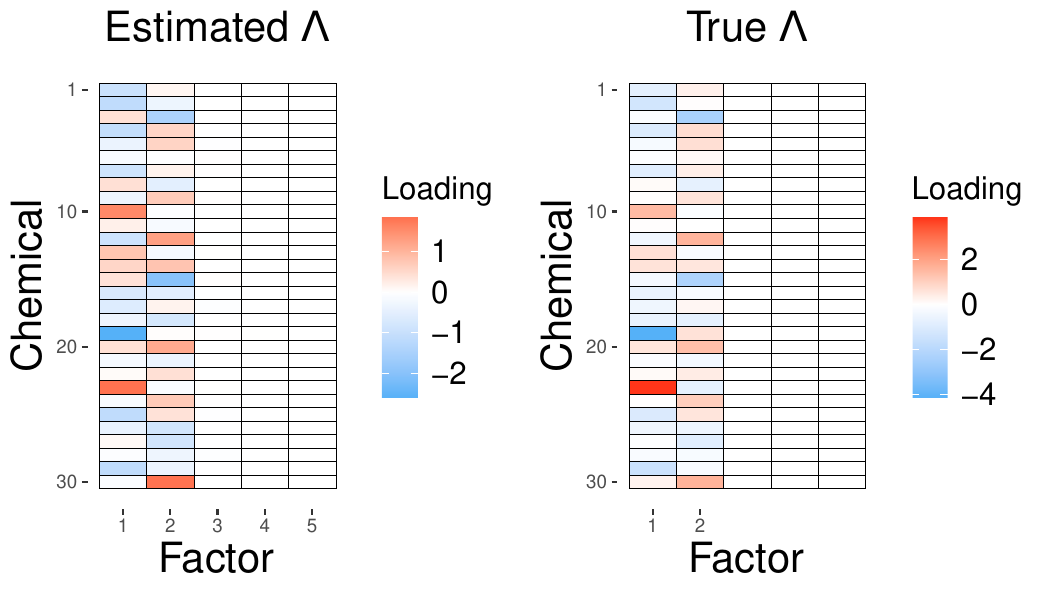}
\caption{The estimated and true entries of loading matrix $\Lambda$ from Simulation 1. Signs in the estimated $\Lambda$ are switched for better visualization.}
\label{Sim1:loadings}
\end{figure}

\begin{figure}[htbp]
\centering
\includegraphics[width=0.7\textwidth]{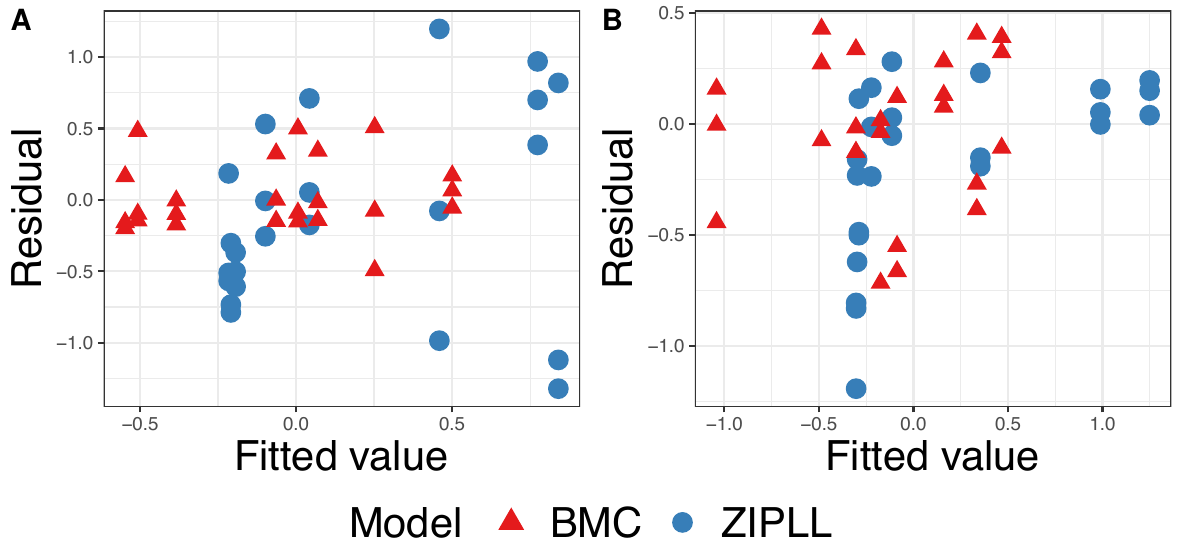}
\caption{Residuals versus fitted values using BMC and ZIPLL in Simulation 1. The residuals and fitted values in \textsf{A} and \textsf{B} are computed using observations and fitted lines from \textsf{A} and \textsf{B} in Figure \ref{Sim1:curves}, respectively. Note that residuals from ZIPLL are obtained by subtracting the fitted values from observations, while those from BMC are the posterior mean of ``normalised'' residuals whose value at $s$th iteration is $(y_{ijk}-\gamma_{ij}^{(s)}f_{ij}^{(s)}(x_{ijk}))/\exp(x_{ijk}\delta^{(s)}_{ij}/2)$.}
\label{Sim1:fitvsres}
\end{figure}

\begin{figure}
\centering
\includegraphics[width = 0.8\textwidth]{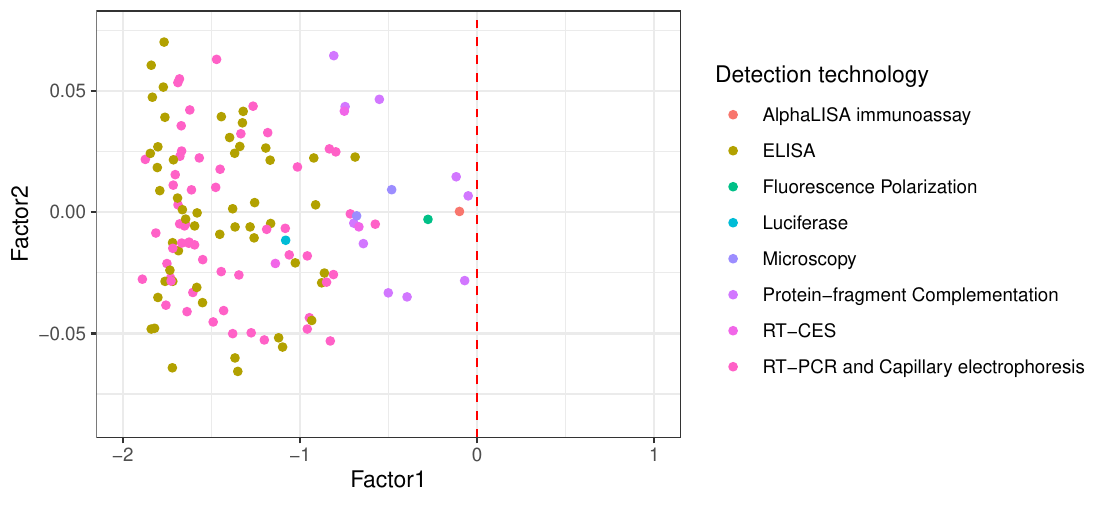}\\
\includegraphics[width = 0.8\textwidth]{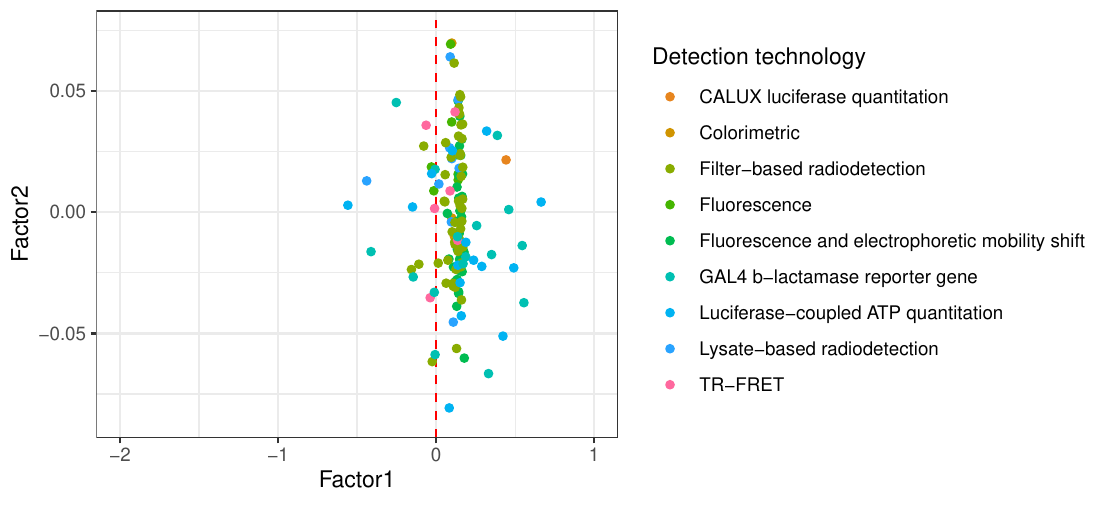}
\caption{First two latent factors for assay endpoints. Each assay endpoint is coloured by its detection technology. Top plot shows assay endpoints with negative loadings, and bottom plot shows those with (mostly) positive loadings on Factor 1. This way, Factor 1 divides detection technologies into two groups.}
\label{obese:latfac}       
\end{figure}
%Viewing Lambda as latent factors and eta as factor loadings 
%Varimax rotation is used to identify latent factors. However, for better visualisation, the original eta is used. Same clustering retained. 

\begin{figure}[htbp]
\centering
\includegraphics[width=0.7\textwidth]{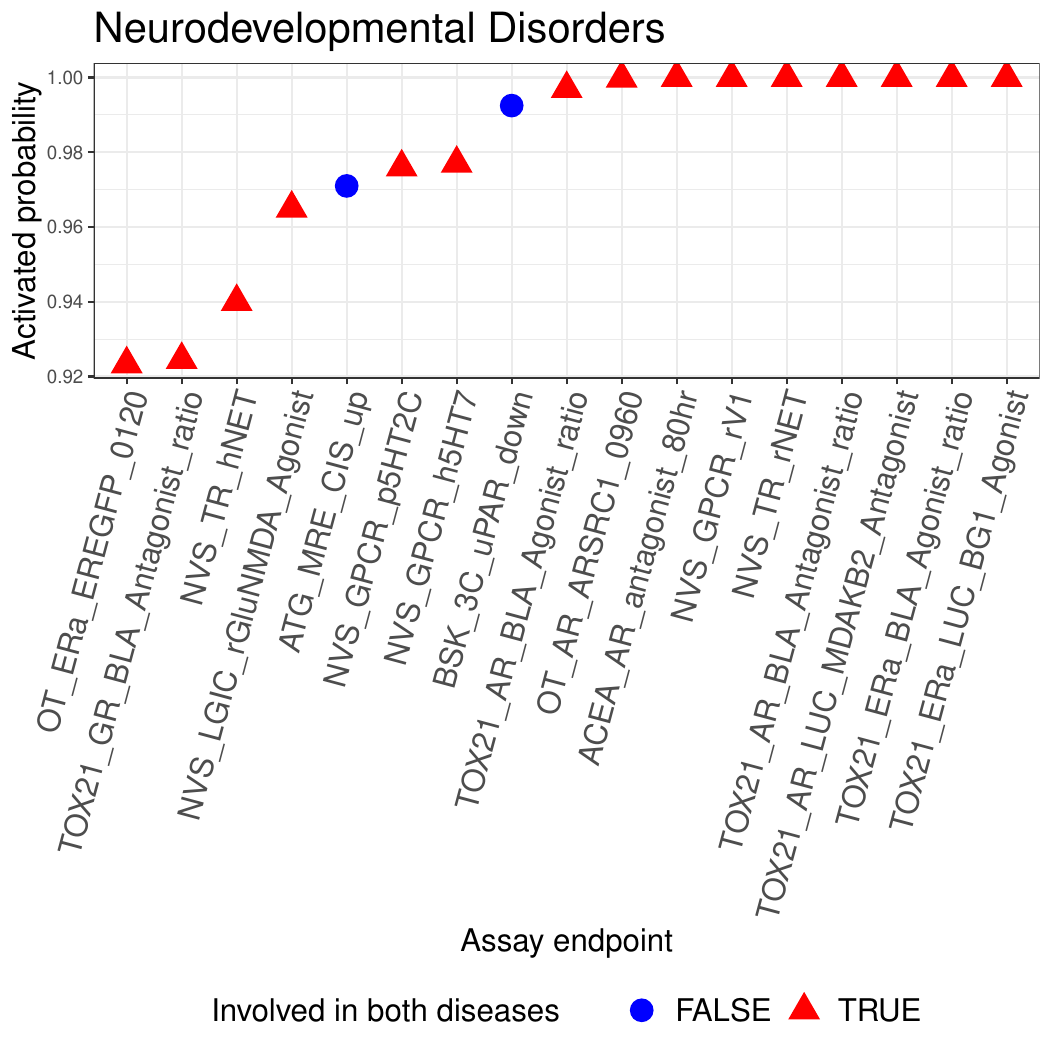}
\caption{Ranks of assay endpoints associated with neurodevelopmental disorders in terms of probabilities to be activated by the top 5 chemicals (Triclosan, p,p'-DDE, BPA, DDT, and 2,4,5-Trichlorophenol). Only a subset of assay endpoints are presented with the activated probabilities higher than 0.9. The assay endpoints with dots are marked uniquely for neurodevelopmental disorders, while those with triangles are marked for two disease classes.}
\label{neuro_disruptassay}
\end{figure}

\begin{figure}[htbp]
\centering
\includegraphics[width=0.8\textwidth]{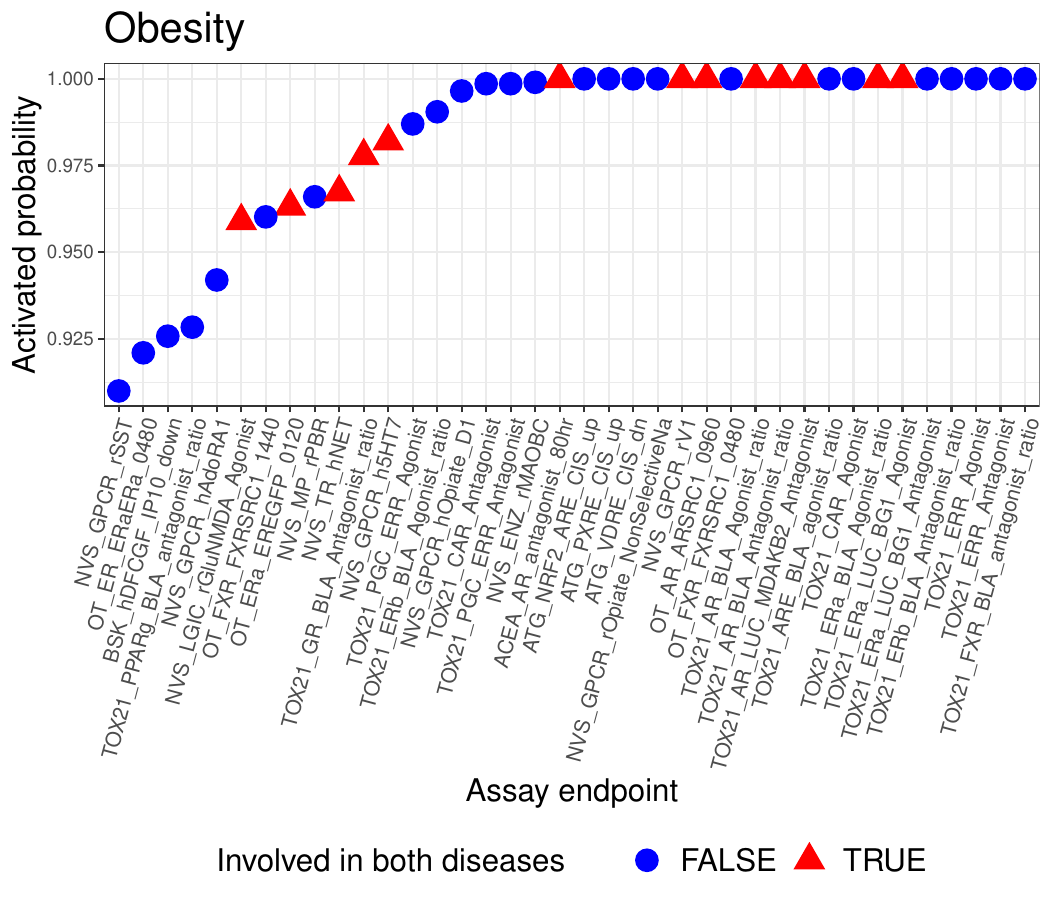}
\caption{Ranks of obesity-related assay endpoints in terms of probabilities to be activated by the top 5 chemicals (Triclosan, p,p'-DDE, BPA, DDT, and 2,4,5-Trichlorophenol). Only a subset of assay endpoints are presented with the activated probabilities higher than 0.9. The assay endpoints with dots are marked uniquely for obesity, while those with triangles are marked for both diseases.}
\label{obese_disruptassay}
\end{figure}

\clearpage
\section{Data} \label{s:data}
This section explains our selection criteria for chemicals and assay endpoints related to neurodevelopmental disorders and obesity in the ToxCast/Tox21 data. The exact procedure to find assay endpoints of interest is as follows: first, molecules are identified that have known associations with each disease through the Comparative Toxicogenomics Database (CTD) \citep{davis2019comparative} and Ingenuity\textsuperscript{\textregistered} PathwayAnalysis (IPA) Knowledgebase (QIAGEN Inc., \url{https://www.qiagenbioinformatics.com/products/ingenuity-pathway-analysis/}). Here, the known associations include the following: molecules are biomarkers of the disease; are known to play a role in the disease etiology; and are therapeutic targets for treatment of the disease. The databases CTD and IPA maintain curated and published associations between molecules and diseases. It is noteworthy that the databases originated from a variety of species and tissue targets. In later steps, the databases are compared to the ToxCast/Tox21 data, whose assay endpoints were derived from a variety of species. Moreover, the assay endpoints in the ToxCast/Tox21 program were tested across several types of tissue targets that may differently mediate the relationship of even the same molecular target and the same assay endpoint. We ensure that a wide variety of species and tissue targets are well represented in both ToxCast/Tox21 and the databases from which molecular targets are identified. It is important to have diverse assay formats for a relevant pathway because it promotes the correct identification of signals over technical errors \citep{huang2014profiling}. Second, we filter molecular targets in ToxCast/Tox21 that overlap with the identified molecules from CTD and IPA. Third, we choose assay endpoints that those overlapping molecular targets are screened over. As a result of these steps, 132 and 352 assay endpoints were identified as relevant to neurodevelopmental disorders and obesity, respectively, which were further filtered based on chemical coverage, as detailed below.

A partial list of chemicals is considered due to a particular interest in human data. We featured a set of overlapping chemicals measured in ToxCast/Tox21 and an existing observational study of environmental risk factors for neurodevelopmental disorders and obesity. In doing so, we believe future application of the ToxCast/Tox21 results to humans will be more viable. A total of 48 chemicals were selected, 30 of which were tested within the above mentioned list of assay endpoints. Due to the reduced list of chemicals, the number of assay endpoints has diminished as well. In addition, \cite{hsieh2015data} found that cytotoxicity is the main confounding factor to activity artefacts. They show that after eliminating signals attributable to cytotoxicity, activity rate could significantly drop. Therefore, following a recommended practice in \cite{judson2016editor}, we retained only the doses lower than a cytotoxicity point for each chemical, which removed two percent of the data. We employed the cytotoxicity median values stored in a variable ``cyto\_pt\_um'' in the \texttt{tcpl} package. Consequently, our final data involve 30 chemicals and 131 and 271 assay endpoints related to neurodevelopmental disorders and obesity, respectively. 

\section{Simulations} \label{s:sims}

For all simulations, we used eight unique doses \{0.301, 0.477, 0.602, 0.845, 1.000, 1.301, 1.602, 2.000\} in log$_{10}~\mu M$ chosen based on the frequency of appearance in the ToxCast/Tox21 data. B-spline knots are set at the minimum value, three quartiles, and the maximum of the doses. 

In Simulation 1, we generated 30 data sets with $K_{ij}=24$ at each combination of chemical and assay endpoint, which represents 3 replicates at each of the eight doses. Elements of $\bm{\eta}$ were drawn independently from the standard normal distribution. Elements in the $m \times q$ matrix $\Lambda$ were sampled as in equations \eqref{mgsp1} and \eqref{mgsp2} with $\nu = 3$, $a_1=2.1$, and $a_2=3.1$, following the note by \cite{durante2017note}. The assay endpoint-specific variances were sampled from $1/\sigma_j^2 \sim Gamma(5/2, 0.5/2)$. For heteroscedastic pairs, $d_{ij}$ was sampled from $N(1.5, 0.1^2)$, which gives $\exp(0.75x_{ijk})\epsilon_{ijk}$ in expectation for the error term. For BMC and ZIPLL, 20,000 samples were drawn, of which 1,000 samples were saved and analysed. First 10,000 samples were discarded as burn-in, and every 10th sample was retained for the next 10,000 samples. Trace plots and effective sample sizes for posterior samples suggested convergence and good mixing. 

In Simulation 2 where misalignment exists between the true data generating process and BMC, we generated data under the ZIPLL model. The number of chemicals $m$ was set to 15, and the number of assay endpoints $J$ to 15. Randomly selected 10\% of the chemical-assay endpoint pairs were held out for prediction of $\gamma_{ij}$. We generated 50 data sets with $K_{ij}=8$ so that each chemical-assay endpoint pair has one observation at each of the eight doses. No replicates at such scarce doses make it impractical to evaluate heteroscedasticity, which is consequently not considered in Simulation 2. The model \[y_{ijk} = \gamma_{ij}f_{ij}(x_{ijk})+\epsilon_{ijk},~\epsilon_{ijk}\sim N(0,0.1^2)\] was considered where around half the pairs were randomly assigned to have the mean effect with $\gamma_{ij} \sim Bernoulli(0.5)$. The dose-response function was 
\[f_{ij}(x_{ijk}) = t_{ij}-\frac{t_{ij}-b_{ij}}{1+\exp\{w_{ij}(\log x_{ijk}-\log a_{ij})\}} \]
where $t_{ij} \sim Unif(0,10)$, $b_{ij}=0$, $a_{ij}=max(x_{ijk})$, and $w_{ij} \sim Unif(1,8).$ 
The RMSE and AUC results are summarised in Table \ref{Simulation2}. From BMC-variants, results from BMC$_0$ are presented only because ZIPLL adopts BMC$_0$ in its $\gamma_{ij}$ testing framework. 

\begin{table}
\caption{Summary of results from Simulation 2. The RMSEs and AUC results of the mean effect probabilities are presented. The displayed values are the mean (standard error) across 50 simulated data sets.}
\label{Simulation2}
\begin{tabular}{c|cccc}
& BMC & BMC$_0$ & ZIPLL  & tcpl \\ \hline
RMSE    & 0.094 (0.002) & 0.094 (0.002) & 0.088 (0.002)  &  0.279 (0.017) \\
In-sample AUC for $\gamma_{ij}$   & 0.982 (0.009) & 0.982 (0.009) & 0.981 (0.009)   &  0.907 (0.018)\\ 
Out-of-sample AUC for $\gamma_{ij}$ & 0.504 (0.109) & 0.504 (0.120) & -   &  - 
\end{tabular}
\end{table}
 
It is remarkable that BMC is able to estimate dose-response trends almost as well as ZIPLL and has similar accuracy in estimating $\gamma_{ij}$ even when ZIPLL is the true data generating process. In Simulation 2, BMC and ZIPLL outperform tcpl models. Smaller RMSEs and higher AUCs from BMC and ZIPLL compared to those from tcpl suggest increased robustness of spline methods than parametric ones for dose-response functions. The improved metrics also indicate benefits of hierarchical methods over the independent curve fitting that ignores correlations between chemicals or assay endpoints. In particular, the achievement of the high in-sample AUC for $\gamma_{ij}$ from BMC is encouraging despite the relatively small number of chemicals and assay endpoints, and model misspecification. Poor predictive AUCs of BMC and BMC$_0$ are expected because the true $\gamma_{ij}$'s are random Bernoulli samples with probability 0.5 without any structure to exploit. 

In multiplicity adjustment simulations (Simulation 3), data are generated from BMC assuming the number of latent factors $q=5$, and hyperparameters to be $\nu = 3$, $a_1=2.1$, and $a_2=3.1$ for the multiplicative gamma process shrinkage prior. Other parameters $\xi = 0.8$, $\bm{\alpha}=(0.3, 1)^T$ are applied, resulting in 20 and 18 true 1's in $\gamma_{ij}$ and in $t_{ij}$, respectively. We fix the true 1's throughout simulations. There are 5 replicates at each of the eight doses, and for heteroscedastic pairs, $\delta_{ij}$ are sampled from $N(1, 0.1^2)$. The assay endpoint-specific variances are given as $1/\sigma_j^2 \sim Gamma(5/2, 0.5/2)$. From 15,000 samples, 1,000 samples were saved and analysed. First 10,000 samples were discarded as burn-in, and every 5th sample was retained for the next 5,000 samples. 

\begin{figure}
    \centering
    \includegraphics[width = 0.8\textwidth]{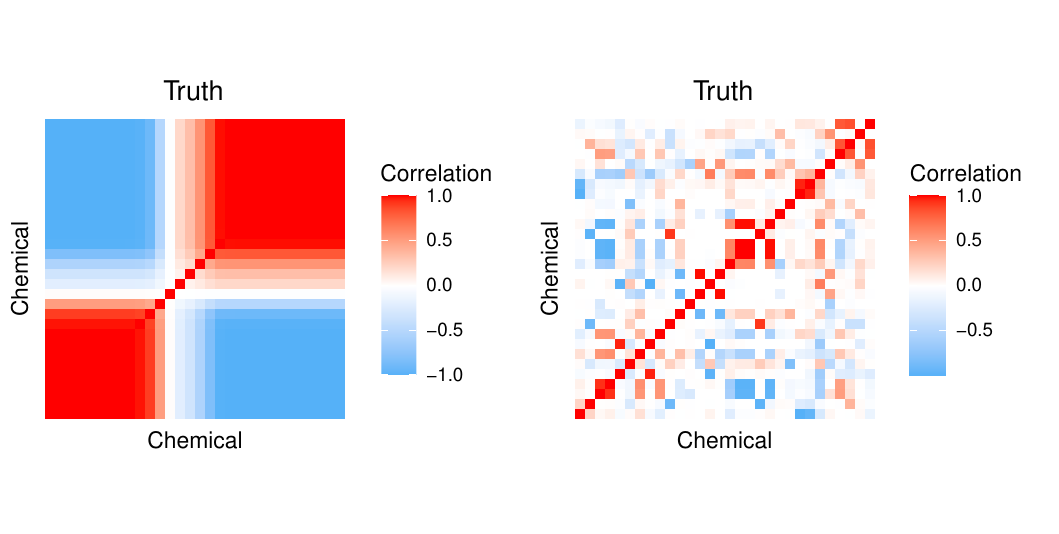}
    \caption{Heat map of correlation matrices among chemicals when they are highly correlated (left) and weakly correlated (right) from Simulation 4.}
    \label{fig:sim7_corr}
\end{figure}

In Simulation 4, We empirically investigate performance of $Pr(\gamma_{ij} = 1 \cup t_{ij} = 1)$ in terms of AUC by varying missingness and correlation structures. The first case is when chemicals are highly correlated. Among $m=30$ chemicals, the correlation structure (see left plot in Figure \ref{fig:sim7_corr}) is given by a latent factor model assuming $q=1$, $\nu=0.1$, $a_1=2.1$, $a_2=3.1$, and $\xi=0.8$. On the other hand, chemicals are weakly correlated in the second case. The correlation matrix (see right plot in Figure \ref{fig:sim7_corr}) is made through a latent factor model assuming $q=10$, $\nu=0.3$, $a_1=2.1$, $a_2=3.1$, and $\xi=0.8$. The number of assay endpoints is $J=10$ with the assay endpoint-specific variances $1/\sigma_j^2$ sampled from $Gamma(5/2, 0.5/2)$. Each pair has $K_{ij}=40$ observations with 5 replicates at each of the eight doses. Missingness varies from 10\% to 50\% in which the maximum level is chosen to reflect ToxCast/Tox21 data. Heteroscedasticity parameters are $\bm{\alpha}=(0.3, 1)^T$ and $\delta_{ij}\sim N(1, 0.1^2)$. Total 10,000 samples were drawn, of which 1,000 samples were saved and analysed. First 5,000 samples were discarded as burn-in, and every 5th sample was retained for the next 5,000 samples. 

\begin{table}
    \caption{Summary of 30 simulation results for $\bm{1}(\gamma_{ij}=1\cup t_{ij}=1)$, $\gamma_{ij}$, and $t_{ij}$ under the highly correlated structure. Different proportions of data are missing: 10\%, 20\%, 30\%, and 50\% missing.}
    \centering
    \begin{tabular}{c|cccc}
            & 10\% & 20\% & 30\% & 50\%   \\ \hline
    In-sample AUC for & \multirow{2}{*}{$>$0.999 ($<$0.001)} & \multirow{2}{*}{$>$0.999 ($<$0.001)} & \multirow{2}{*}{$>$0.999 ($<$0.001)} & \multirow{2}{*}{$>$0.999 ($<$0.001)}\\
    $\bm{1}(\gamma_{ij} = 1 \cup t_{ij} = 1)$ & & & \\
    Out-of-sample AUC for & \multirow{2}{*}{0.966 (0.035)} & \multirow{2}{*}{0.965 (0.022)} & \multirow{2}{*}{0.958 (0.024)} & \multirow{2}{*}{0.915 (0.052)} \\
    $\bm{1}(\gamma_{ij} = 1 \cup t_{ij} = 1)$ & & &\\
    In-sample AUC for $\gamma_{ij}$ & $>$0.999 ($<$0.001) & $>$0.999 ($<$0.001) & $>$0.999 ($<$0.001) & 0.999 (0.001) \\
    Out-of-sample AUC for $\gamma_{ij}$ & 0.954 (0.037) & 0.952 (0.031) & 0.947 (0.031) & 0.906 (0.048) \\
    In-sample AUC for $t_{ij}$ & 0.999 (0.001) & 0.999 (0.001) & 0.998 (0.002) & 0.994 (0.006)\\
    Out-of-sample AUC for $t_{ij}$ & 0.961 (0.028) & 0.955 (0.027) & 0.947 (0.027) & 0.898 (0.056) % \\
    \end{tabular}
    \label{tab:missing1}
\end{table}    
    
\begin{table}
    \caption{Summary of 30 simulation results for $\bm{1}(\gamma_{ij}=1\cup t_{ij}=1)$, $\gamma_{ij}$, and $t_{ij}$ under the weakly correlated structure. Different proportions of data are missing: 10\%, 20\%, 30\%, and 50\% missing.}
    \centering
    \begin{tabular}{c|cccc}
            & 10\% & 20\% & 30\% & 50\%   \\ \hline
    In-sample AUC for & \multirow{2}{*}{$>$0.999 ($<$0.001)} & \multirow{2}{*}{$>$0.999 ($<$0.001)} & \multirow{2}{*}{$>$0.999 ($<$0.001)} & \multirow{2}{*}{$>$0.999 ($<$0.001)} \\
    $\bm{1}(\gamma_{ij} = 1 \cup t_{ij} = 1)$ & & & \\
    Out-of-sample AUC for & \multirow{2}{*}{0.800 (0.101)} & \multirow{2}{*}{0.757 (0.066)} & \multirow{2}{*}{0.712 (0.071)} & \multirow{2}{*}{0.573 (0.058)}\\
    $\bm{1}(\gamma_{ij} = 1 \cup t_{ij} = 1)$ & & & \\
    In-sample AUC for $\gamma_{ij}$ & $>$0.999 ($<$0.001) & $>$0.999 ($<$0.001) & 0.999 (0.001) & 0.999 (0.002) \\
    Out-of-sample AUC for $\gamma_{ij}$ & 0.805 (0.088) & 0.762 (0.066) & 0.712 (0.069) & 0.578 (0.059) \\
    In-sample AUC for $t_{ij}$ & 0.999 ($<$0.001) & 0.998 (0.001) & 0.996 (0.003) & 0.992 (0.005)\\
    Out-of-sample AUC for $t_{ij}$ & 0.811 (0.085) & 0.762 (0.069) & 0.716 (0.067) & 0.548 (0.049) %\\
    \end{tabular}
    \label{tab:missing2}
\end{table}

In Tables \ref{tab:missing1} and \ref{tab:missing2}, in-sample AUCs perform well and do not seem much affected by missingness or different correlation structures. The in-sample AUCs are above 0.990 for any indicators across different rates of missingness with or without a strong correlation structure. Out-of-sample AUCs, however, exhibit a clearer decline as missingness increases. Despite the decreasing trend, the out-of-sample AUCs remain high at around 0.9 even with 50\% missingness when chemicals are highly correlated and the model is well-specified. This is due to the fact that BMC properly exploits the correlated structure of chemicals through a latent factor model. However, performance can decline in weak correlation cases as borrowing of information pays less dividends. In Table \ref{tab:missing2}, out-of-sample AUCs are acceptable ($>0.7$) for up to 30\% missingness in all activity metrics, whereas they are only slightly better than random guessing when half the data are missing. 

\section{ToxCast/Tox21 Prediction Results} \label{s:prediction}

\begin{figure}[htbp]
\centering
\includegraphics[width=0.8\textwidth]{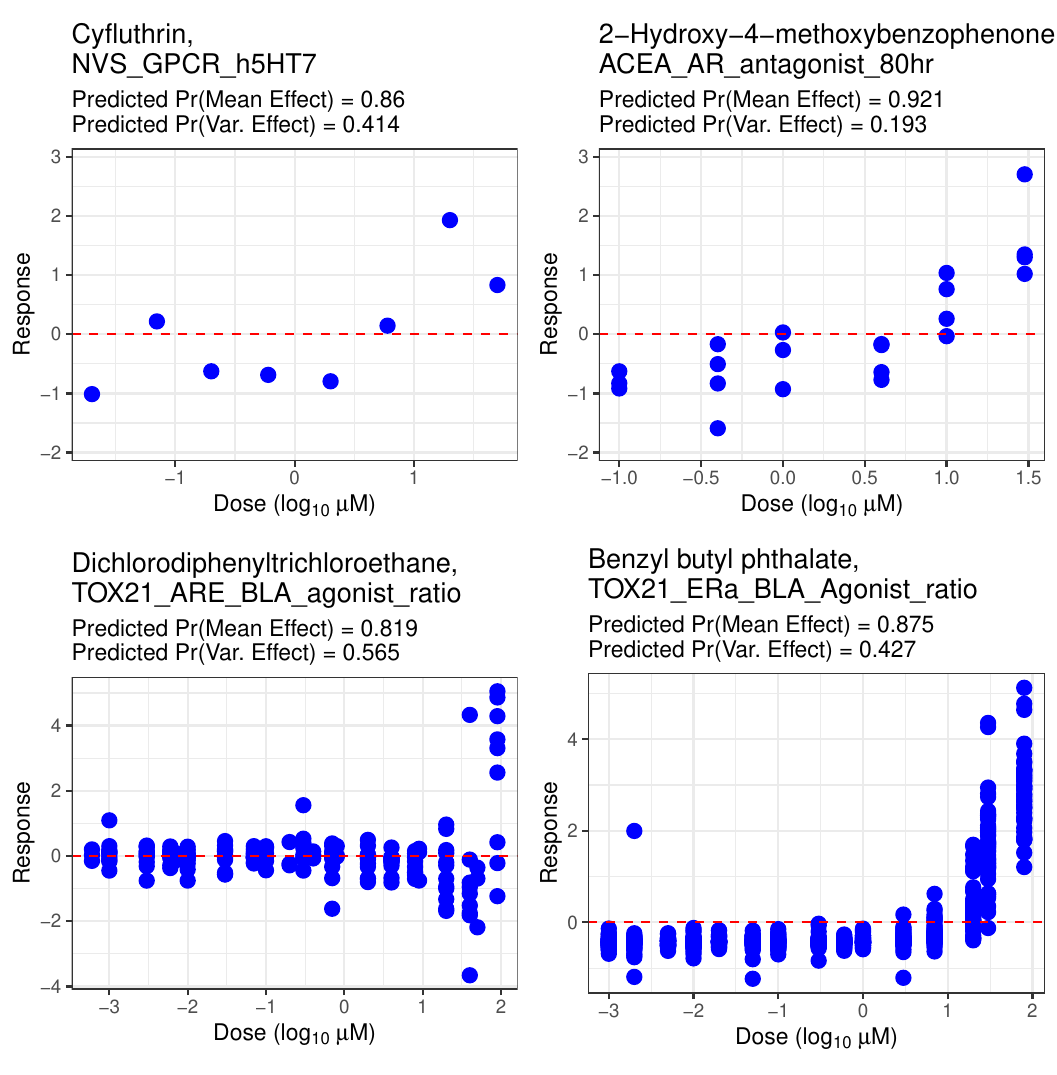}
\caption{Results for select chemical-assay endpoint pairs with BMC's posterior probabilities for mean effects and heteroscedastic effects. Blue points are observations of the held-out pairs, and a red dashed line is a horizontal line at 0 for reference.}
\label{pred_active}
\end{figure}

Figures \ref{pred_active} shows select predicted results for held-out pairs in the ToxCast/Tox21 data. BMC offers probabilities of the mean and the variance effect for activity profiles which are often the primary focus of many studies and help researchers to prioritise chemicals for further testing. As discussed in Section \ref{ToxCast}, $\xi$ is estimated to be $1.147$ and $\alpha_0$ to be $-0.208$, yielding $\Phi(\hat{\xi}) = 0.874$ and $\Phi(\hat{\alpha_0}) = 0.418$. We found that cutoffs chosen as values of the standard normal CDF at the global parameter estimates provide reasonable $\{0,1\}$ classification from the posterior probabilities. Based on the cutoffs of 0.874 for ``Pr(Mean Effect)'' and 0.418 for ``Pr(Var Effect)'', we conclude that the top-left pair is not active, the top-right is disrupted in the mean only, the bottom-left has the variance effect alone without mean effects, and the bottom-right has both mean and variance effects, which coincides with visual judgement. 

\section{Posterior Computation} \label{s:posteriors}
Under the prior specification in section \ref{Model}, posterior samples are obtained by iterating the following partially collapsed Markov chain Monte Carlo (MCMC) sampler. 
\begin{enumerate}
    \item[] \textbf{\textit{Heteroscedasticity}}
    \item Given that $\bm{\alpha} = (\alpha_0, \alpha_1)^T \sim N(\bm{\mu}_{\alpha}, V_{\alpha})$ \textit{a priori}, update $\bm{\alpha}$ from 
        \begin{align*}
            (\bm{\alpha}|U, W) \sim N(V^*_{\alpha}\bm{\mu}^*_{\alpha}, V^*_{\alpha}) \text{ with }\\
            V^*_{\alpha} = (V_{\alpha}^{-1} + W^TW)^{-1} \text{ and } \bm{\mu}^*_{\alpha} = V_{\alpha}^{-1}\bm{\mu}_{\alpha}+W^T\text{vec}(U)
        \end{align*} 
        where $U$ is a $m\times J$ matrix whose $(i,j)$ element $u_{ij}$, and $W = \begin{bmatrix} \bm{1} & \text{vec}(M) \end{bmatrix}$ is a $mJ\times 2$ matrix. The second column $\text{vec}(M)$ is the vectorisation of $M$ which is a $m\times J$ matrix with $\bm{\lambda}_i^T\bm{\eta}_j$ for its $(i,j)$ element. 
        
    \item Update $u_{ij}$ from 
    \begin{align*}
        (u_{ij}|t_{ij} = 1, \bm{\alpha}, (\bm{\lambda}_i^T\bm{\eta}_j)) \sim TN_{(0, \infty)}((\alpha_0 + (\bm{\lambda}_i^T\bm{\eta}_j)\alpha_1), 1), \\
        (u_{ij}|t_{ij} = 0, \bm{\alpha},(\bm{\lambda}_i^T\bm{\eta}_j)) \sim TN_{(-\infty, 0)}((\alpha_0 + (\bm{\lambda}_i^T\bm{\eta}_j)\alpha_1), 1).
    \end{align*}
        
    \item Update $t_{ij}$ and $\delta_{ij}$ simultaneously using the Metropolis algorithm. Propose $t^p_{ij}$ as follows: for each $j$, choose random number of elements and random indices to update. For those selected $(i,j)$ pairs, flip zero and one. Given the proposed $t^p_{ij},$ propose $\delta^p_{ij}$ using t-distribution with 4 degrees of freedom centered at the current $\delta^c_{ij}$. Accept $(t^p_{ij}, \delta^p_{ij})$ with probability $\min\{1,r\}$ in which
    \begin{align*}
        r =& \frac{\prod_{k=1}^{K_{ij}}N(y_{ijk}-\gamma_{ij}(\vt{x}^B_{ijk})^T\bm{\beta}_{ij}; 0, \exp(x_{ijk}\delta^p_{ij}/2)^2\sigma^2_j)}{\prod_{k=1}^{K_{ij}}N(y_{ijk}-\gamma_{ij}(\vt{x}^B_{ijk})^T\bm{\beta}_{ij}; 0, \exp(x_{ijk}\delta^c_{ij}/2)^2\sigma^2_j)} \times \\ &\frac{Bernoulli(t^p_{ij};\Phi(\alpha_0+(\bm{\lambda}_i^T\bm{\eta}_j)\alpha_1))\{N(\delta^p_{ij}|t^p_{ij}=1;0,v_{\delta})\vt{1}(t^p_{ij}=1)+1\times \vt{1}(t^p_{ij}=0)\}}{Bernoulli(t^c_{ij};\Phi(\alpha_0+(\bm{\lambda}_i^T\bm{\eta}_j)\alpha_1))\{N(\delta^c_{ij}|t^c_{ij}=1;0,v_{\delta})\vt{1}(t^c_{ij}=1)+1\times \vt{1}(t^c_{ij}=0)\}}
    \end{align*}
    and $(\vt{x}^B_{ijk})^T$ is the $k$th row of the B-spline basis matrix $X_{ij}$. 
    \item[]
    \item[] \textbf{\textit{Functional Mean}} \\
    Once Steps 1-3 are completed in every iteration, the data $(X,Y)$ need to be reformulated: $y_{ijk}$ is replaced by $y_{ijk}/\exp(x_{ijk}\delta_{ij}/2)$, and $\vt{x}_{ijk}^B$ by $\vt{x}_{ijk}^B/\exp(x_{ijk}\delta_{ij}/2)$.
    
    \item Update $\bm{\lambda}_i$ from 
    \begin{align*}
        &(\bm{\lambda}_i|\bm{\eta}, \bm{z}_i, \xi,\bm{\alpha},\bm{u}_i) \sim \\ &N_q((D_i^{-1}+(1+\alpha_1^2)\bm{\eta}^T\bm{\eta})^{-1}
        (\bm{\eta}^T\tilde{\bm{z}}_i + \alpha_1\bm{\eta}^T\bm{u}_i-\alpha_0\alpha_1\bm{\eta}^T\bm{1}_J), (D_i^{-1}+(1+\alpha_1^2)\bm{\eta}^T\bm{\eta})^{-1}) 
    \end{align*}
    where $\bm{\eta}$ is a $J \times q$ matrix whose $j$th row is $\bm{\eta}_j^T$, $D_i^{-1}$ = diag($\phi_{i1}\tau_1, \cdots, \phi_{iq}\tau_q$), $\tilde{\bm{z}}_i = (z_{i1}-\xi,\cdots,z_{iJ}-\xi)^T$, and $\bm{u}_i = (u_{i1},\cdots,u_{iJ})^T$. 

    \item Update $\bm{\eta}_j$ from
    \begin{align*}
        &(\bm{\eta}_j|\Lambda, \bm{z}_j, \xi, \bm{\alpha}, \bm{u}_j) \sim \\  &N_q((I_q + (1+\alpha_1^2)\Lambda^T\Lambda)^{-1}(\Lambda^T\tilde{\bm{z}}_j + \alpha_1\Lambda^T\bm{u}_j-\alpha_0\alpha_1\Lambda^T\bm{1}_m), (I_q + (1+\alpha_1^2)\Lambda^T\Lambda)^{-1})
    \end{align*}
    where $\Lambda$ is a $m \times q$ matrix whose $i$th row is $\bm{\lambda}^T_i$, $\tilde{\bm{z}}_j = (z_{1j}-\xi, \cdots, z_{mj}-\xi)^T$, and $\bm{u}_j = (u_{1j}, \cdots, u_{mj})^T$.

    \item Given that $\xi \sim N(\mu_{\xi}, \sigma^2_{\xi})$ \textit{a priori}, update $\xi$ from 
    \begin{align*}
        (\xi|Z,\Lambda, \bm{\eta}) \sim N\left((1/\sigma_{\xi}^2+\sum_{j=1}^Jm_j)^{-1}\left\{\mu_{\xi}/\sigma_{\xi}^2+\sum_{j=1}^J\sum_{i=1}^{m_j}(z_{ij}-\bm{\lambda}_i^T\bm{\eta}_j)\right\}, (1/\sigma_{\xi}^2+\sum_{j=1}^Jm_j)^{-1}\right).
    \end{align*}
        
    \item Update $z_{ij}$ from 
    \begin{align*}
        (z_{ij}|\gamma_{ij}=1, \bm{\lambda}_i, \bm{\eta}_j, \xi) &\sim TN_{(0,\infty)}(\xi+\bm{\lambda}_i^T \bm{\eta}_j,1), \\
        (z_{ij}|\gamma_{ij}=0, \bm{\lambda}_i, \bm{\eta}_j, \xi) &\sim TN_{(-\infty,0)}(\xi+\bm{\lambda}_i^T \bm{\eta}_j,1)
    \end{align*}
    where $TN_{(a,b)}(\mu,\sigma^2)$ denotes a normal distribution truncated to the interval $(a,b)$ with mean $\mu$, variance $\sigma^2$.
    
    \item Update $\gamma_{ij}$ from the conditional Bernoulli distribution with $\bm{\beta}_{ij}$ marginalised out. With $\pi_{ij} = \Phi(\xi+\bm{\lambda}_i^T \bm{\eta}_j)$,
    \begin{align} 
        &Pr(\gamma_{ij}=1|\vt{y}_{ij}, X_{ij}, \sigma_j^2, \Sigma_j, \pi_{ij}) \nonumber\\
        &\propto  \pi_{ij}|\Sigma_jX_{ij}^TX_{ij}/\sigma_j^2+I_p|^{-1/2} \times\exp\left(\frac{1}{2\sigma_j^4}\vt{y}_{ij}^TX_{ij}\left(X_{ij}^TX_{ij}/\sigma_j^2+\Sigma_j^{-1}\right)^{-1} X^T_{ij}\vt{y}_{ij}\right), \label{gam1nobeta}\\ 
        &Pr(\gamma_{ij}=0|\vt{y}_{ij}, X_{ij}, \sigma_j^2, \Sigma_j, \pi_{ij}) \propto  (1-\pi_{ij}), \label{gam0nobeta}
    \end{align}
    \[(\gamma_{ij}|\vt{y}_{ij}, X_{ij}, \sigma_j^2, \Sigma_j, \pi_{ij}) \sim Bernoulli\left(\frac{(\ref{gam1nobeta})}{(\ref{gam1nobeta})+(\ref{gam0nobeta})}\right) \] where $\vt{y}_{ij}=[y_{ij,1},\dots,y_{ij,K_{ij}}]^T$. 
        
    \item Update $\phi_{il}$ and $\zeta_h$ as in \cite{bhattacharya2011sparse}. Hyperparameter selection and posterior distributions are fully explained in \cite{durante2017note} and \cite{bhattacharya2011sparse}. We shall not repeat the sampling algorithms here. 
    
    \item Update $\bm{\beta}_{ij}$ from the conditional normal distribution only if $\gamma_{ij}=1$
    \begin{align*}
        &(\bm{\beta}_{ij}|\gamma_{ij}=1,\vt{y}_{ij}, X_{ij}, \sigma_j^2, \Sigma_j) \\&\sim N_p\left(\left(\Sigma_j^{-1}+X^T_{ij}X_{ij}/\sigma_j^2\right)^{-1}X^T_{ij}\vt{y}_{ij}/\sigma_j^2,\left(\Sigma_j^{-1}+X^T_{ij}X_{ij}/\sigma_j^2\right)^{-1}\right).
    \end{align*}
    
    \item Update $\Sigma_j$ from 
    \[\left(\Sigma_j^{-1}|\bm{\beta}_{1j},\dots,\bm{\beta}_{m_j,j}\right) \sim Wish\left(a+m_j, \left(R+\sum_{i=1}^{m_j}\bm{\beta}_{ij}\bm{\beta}_{ij}^T\right)^{-1}\right). \]
    
    \item[] \textbf{\textit{Assay endpoint-specific variance}}
    \item Update $\sigma^2_j$ from 
    \begin{align*}
        &(1/\sigma_j^2|\vt{y}_{ij}, \gamma_{ij}, X_{ij}, \bm{\beta}_{ij} \forall i=1,\dots,m_j) \\
        &\sim Gamma\left(\frac{\nu_0+\sum_{i=1}^{m_j}K_{ij}}{2}, \frac{\nu_0\sigma_0^2+\sum_{i=1}^{m_j}\sum_{k=1}^{K_{ij}}(y_{ijk}-\gamma_{ij}(\vt{x}_{ijk}^B)^T\bm{\beta}_{ij})^2}{2}\right).
    \end{align*}
\end{enumerate}

The above algorithm can be easily modified to a logistic model via P\'{o}lya-Gamma augmentation \citep{polson2013bayesian}. 

\end{document}